\documentclass[fleqn,usenatbib]{config/mnras}

\usepackage{newtxtext,newtxmath}
\usepackage{anyfontsize}

\usepackage[T1]{fontenc}
\usepackage{multicol}
\usepackage{paralist}
\usepackage{subcaption}
\usepackage[dvipsnames]{xcolor}

\DeclareRobustCommand{\VAN}[3]{#2}
\let\VANthebibliography\thebibliography
\def\thebibliography{\DeclareRobustCommand{\VAN}[3]{##3}\VANthebibliography}

\usepackage{comment}
\usepackage{graphicx}	
\usepackage{amsmath}	
\usepackage{placeins}	

\usepackage{etoolbox}
\makeatletter
\newcommand\sendemail[4]{
\edef\@tempa{mailto:#1?subject=#2&body=#3 }%
\edef\@tempb{\expandafter\html@spaces\@tempa\@empty}%
\href{\@tempb}{#4}}

\catcode\%=11
\def\html@spaces#1 #2{#1
\catcode\%=14
\makeatother

\usepackage{adjustbox}
\usepackage{textgreek}
\usepackage{xargs}
\usepackage{xspace}

\newcommand{\orcid}[2]{\href{http://orcid.org/#2}{#1}}
\newcommand{\orcidsymb}[2]{\href{http://orcid.org/#2}{#1\adjustbox{trim={-.15\width} {0\height} {-.15\width} {0\height},clip}{\includegraphics[height=10pt]{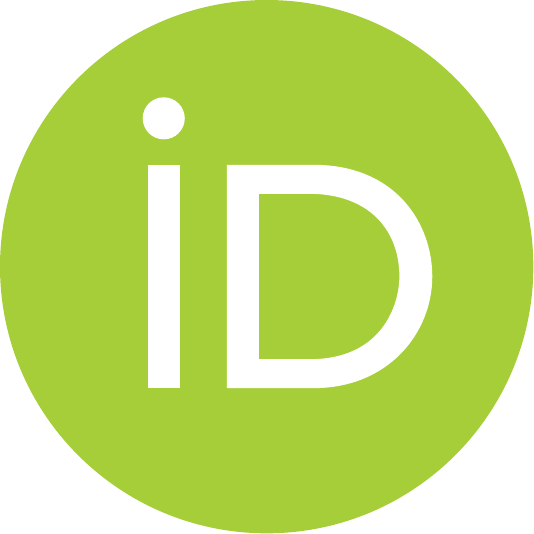}}}}

\newcommand{\citationneeded}{\textcolor{ForestGreen}{$^{\rm citation\;needed}$}}
\let\oldtextsigma\textsigma
\renewcommand{\textsigma}{\oldtextsigma\xspace}
\let\oldtextdegree\textdegree
\renewcommand{\textdegree}{\oldtextdegree\xspace}

\newcommand{\kms}{\ensuremath{\mathrm{km\,s^{-1}}}\xspace}
\newcommand{\Msun}{\ensuremath{{\rm M}_\odot}\xspace}
\newcommand{\Zsun}{\ensuremath{{\rm Z}_\odot}\xspace}
\newcommand{\yr}{\ensuremath{{\rm yr}}\xspace}
\newcommand{\Myr}{\ensuremath{{\rm Myr}}\xspace}
\newcommand{\Gyr}{\ensuremath{{\rm Gyr}}\xspace}
\newcommand{\peryr}{\ensuremath{{\rm yr^{-1}}}\xspace}
\newcommand{\Lsun}{\hbox{\,${\rm L}_\odot$}}
\newcommand{\kpc}{\text{kpc}\xspace}
\newcommand{\ZH}{\text{[Z/H]}\xspace}
\newcommand{\MUV}{\ensuremath{M_\mathrm{UV}}\xspace}
\newcommandx{\percm}[1][1=3]{\ensuremath{\mathrm{cm}^{-#1}}\xspace}	

\newcommandx{\lambdar}[2][1=R,2=]{\ensuremath{\lambda_{\rm {#1}}{#2}}\xspace}
\newcommand{\eps}{\ensuremath{\epsilon}\xspace}
\newcommand{\mstar}{\ensuremath{M_\star}\xspace}
\newcommand{\mdyn}{\ensuremath{M_\mathrm{dyn}}\xspace}
\newcommand{\re}{\ensuremath{R_\mathrm{e}}\xspace}
\newcommand{\vstar}{\ensuremath{v_\star}\xspace}
\newcommand{\vnai}{\ensuremath{v_{\NaI}}\xspace}
\newcommand{\sigmastar}{\ensuremath{\sigma_\star}\xspace}
\newcommand{\sigmaestar}{\ensuremath{\sigma_{\star,\mathrm{e}}}\xspace}
\newcommand{\vperc}[1]{\ensuremath{v_{#1}}\xspace}

\newcommand{\vesc}{\ensuremath{v_\mathrm{esc}}\xspace}
\newcommand{\nelec}{\ensuremath{n_\mathrm{e}}\xspace}
\newcommand{\Telec}{\ensuremath{T_\mathrm{e}}\xspace}
\newcommand{\Rout}{\ensuremath{R_\mathrm{out}}\xspace}
\newcommand{\vout}{\ensuremath{v_\mathrm{out}}\xspace}
\newcommandx{\Mout}[2][1=,2=]{\ensuremath{M_{\mathrm{out}{#2}}^{#1}}\xspace}
\newcommandx{\Mdotout}[2][1=,2=]{\ensuremath{\dot{M}_{\mathrm{out}{#2}}^{#1}}\xspace}
\newcommand\sbullet[1][.5]{\mathbin{\vcenter{\hbox{\scalebox{#1}{$\bullet$}}}}}
\newcommand{\mbh}{\ensuremath{M_{\sbullet[0.85]}}\xspace}

\newcommandx{\fluxdcgs}[1][1=-20]{\ensuremath{\times 10^{#1}~\mathrm{erg\,s^{-1}\,cm^{-2}\,\AA^{-1}}}\xspace}
\newcommandx{\fluxcgs}[3][1=-20,2=\times,3=]{\ensuremath{{#2}10^{#1}~\mathrm{erg\,s^{{#3}-{#3}1}\,cm^{{#3}-{#3}2}}}\xspace}
\newcommandx{\powercgs}[2][1=44,2=\times]{\ensuremath{{#2}10^{#1}~\mathrm{erg\,s^{-1}}}\xspace}
\newcommandx{\ergs}{\ensuremath{\mathrm{erg\,s^{-1}}}\xspace}
\newcommand{\AV}{\ensuremath{A_V}\xspace}

\newcommand{\cigale}{{\sc cigale}\xspace}
\newcommand{\jwst}{\textit{JWST}\xspace}
\newcommand{\hst}{\textit{HST}\xspace}
\newcommand{\ppxf}{{\sc ppxf}\xspace}
\newcommand{\prospector}{{\sc prospector}\xspace}
\newcommand{\emcee}{{\sc emcee}\xspace}
\newcommand{\eazy}{{\sc eazy}\xspace}
\newcommand{\cloudy}{{\sc cloudy}\xspace}
\newcommand{\pyneb}{{\sc pyneb}\xspace}
\newcommandx{\mappings}[1][1=]{{\sc mappings{#1}}\xspace}
\newcommand{\galfit}{{\sc galfit}\xspace}
\newcommand{\qubespec}{{\sc qubespec}\xspace}
\newcommand{\pysersic}{{\sc pysersic}\xspace}
\newcommand{\mpt}{{\sc mpt}\xspace}

\newcommand{\blackthunder}{BlackTHUNDER\xspace}
\newcommand{\Mdynvalue}{$\Mdyn = 2.0\pm0.5 \times 10^{11}$~\MSun}

\defcitealias{gordon+2003}{G03}

\usepackage{amsmath}	
\usepackage{textgreek}
\usepackage{xargs}
\usepackage{xspace}

\let\oldAA\AA
\renewcommand{\AA}{\text{\oldAA}\xspace}
\newcommand{\mum}{\text{\textmu m}\xspace}

\newcommand{\Lyalpha}{\text{Ly\,\textalpha}\xspace}
\newcommand{\Lybeta}{\text{Ly\,\textbeta}\xspace}
\newcommand{\Halpha}{\text{H\,\textalpha}\xspace}
\newcommand{\Hbeta}{\text{H\,\textbeta}\xspace}
\newcommand{\Hgamma}{\text{H\,\textgamma}\xspace}
\newcommand{\Hdelta}{\text{H\,\textdelta}\xspace}
\newcommand{\Paalpha}{\text{Pa\,\textalpha}\xspace}
\newcommand{\Pabeta}{\text{Pa\,\textbeta}\xspace}
\newcommand{\Pagamma}{\text{Pa\,\textgamma}\xspace}
\newcommand{\Padelta}{\text{Pa\,\textdelta}\xspace}
\newcommand{\Hepsilon}{\text{H\,\textepsilon}\xspace}

\newcommandx{\permittedEL}[6][1=O,2=III,3=,4=,5=,6=]{\text{{#1}\,{\sc {#2}}{#3}{#4}{#5}{#6}}\xspace}
\newcommandx{\semiforbiddenEL}[6][1=O,2=III,3=,4=,5=,6=]{\text{{#1}\,{\sc{#2}}]{#3}{#4}{#5}{#6}}\xspace}
\newcommandx{\forbiddenEL}[6][1=O,2=III,3=,4=,5=,6=]{\text{[{#1}\,{\sc{#2}}]{#3}{#4}{#5}{#6}}\xspace}

\newcommand{\HI}{\permittedEL[H][i]}
\newcommand{\HII}{\permittedEL[H][ii]}

\newcommand{\NV}{\permittedEL[N][v]}
\newcommandx{\NVL}[1][1=1243]{\permittedEL[N][v][\textlambda][#1]}
\newcommandx{\NVall}{\permittedEL[N][v][\textlambda][\textlambda][1239,][1243]}

\newcommandx{\CIIall}{\semiforbiddenEL[C][ii][\textlambda][\textlambda][2324--][2329]}

\newcommand{\NIV}{\semiforbiddenEL[N][iv]}
\newcommandx{\NIVL}[1][1=1486]{\semiforbiddenEL[N][iv][\textlambda][#1]}

\newcommand{\CIV}{\permittedEL[C][iv]}
\newcommandx{\CIVL}[1][1=1550]{\permittedEL[C][iv][\textlambda][#1]}
\newcommand{\CIVall}{\permittedEL[C][iv][\textlambda][\textlambda][1548,][1551]}

\newcommand{\HeII}{\permittedEL[He][ii]}
\newcommandx{\HeIIL}[1][1=1640]{\permittedEL[He][ii][\textlambda][#1]}

\newcommand{\semiOIII}{\semiforbiddenEL[O][iii]}
\newcommandx{\semiOIIIL}[1][1=1666]{\semiforbiddenEL[O][iii][\textlambda][#1]}
\newcommand{\semiOIIIall}{\semiforbiddenEL[O][iii][\textlambda][\textlambda][1661,][1666]}

\newcommand{\NIII}{\semiforbiddenEL[N][iii]}
\newcommandx{\NIIIL}[1][1=1750]{\semiforbiddenEL[N][iii][\textlambda][#1]}
\newcommand{\NIIIall}{\semiforbiddenEL[N][iii][\textlambda][\textlambda][1747--][1754]}

\newcommandx{\CIII}{\semiforbiddenEL[C][iii]}
\newcommandx{\CIIIL}[1][1=1909]{\semiforbiddenEL[C][iii][\textlambda][#1]}
\newcommand{\CIIIall}{\semiforbiddenEL[C][iii][\textlambda][\textlambda][1907,][1909]}

\newcommand{\NeIV}{\forbiddenEL[Ne][iv]}
\newcommandx{\NeIVL}[1][1=2424]{\forbiddenEL[Ne][iv][\textlambda][#1]}
\newcommand{\NeIVall}{\forbiddenEL[Ne][iv][\textlambda][\textlambda][2422,][2424]}

\newcommand{\MgII}{\permittedEL[Mg][ii]}
\newcommandx{\MgIIL}[1][1=2803]{\permittedEL[Mg][ii][\textlambda][#1]}
\newcommand{\MgIIall}{\permittedEL[Mg][ii][\textlambda][\textlambda][2796,][2803]}

\newcommand{\NeV}{\forbiddenEL[Ne][v]}
\newcommandx{\NeVL}[1][1=3426]{\forbiddenEL[Ne][v][\textlambda][#1]}
\newcommand{\NeVall}{\forbiddenEL[Ne][v][\textlambda][\textlambda][3346,][3426]}

\newcommand{\OIIperm}{\permittedEL[O][ii]}
\newcommand{\OII}{\forbiddenEL[O][ii]}
\newcommandx{\OIIL}[1][1=3726]{\forbiddenEL[O][ii][\textlambda][#1]}
\newcommandx{\OIIall}[2][1={3726,},2=3729]{\forbiddenEL[O][ii][\textlambda][\textlambda][#1][#2]}

\newcommand{\NeIII}{\forbiddenEL[Ne][iii]}
\newcommandx{\NeIIIL}[1][1=3869]{\forbiddenEL[Ne][iii][\textlambda][#1]}
\newcommand{\NeIIIall}{\forbiddenEL[Ne][iii][\textlambda][\textlambda][3869,][3967]}

\newcommand{\OIII}{\forbiddenEL[O][iii]}
\newcommandx{\OIIIL}[1][1=5007]{\forbiddenEL[O][iii][\textlambda][#1]}
\newcommand{\OIIIall}{\forbiddenEL[O][iii][\textlambda][\textlambda][4959,][5007]}

\newcommandx{\NIL}[1][1=5200]{\forbiddenEL[N][i][\textlambda][#1]}
\newcommand{\NIall}{\forbiddenEL[N][i][\textlambda][\textlambda][5198,][5200]}

\newcommand{\OI}{\forbiddenEL[O][i]}
\newcommandx{\OIL}[1][1=6300]{\forbiddenEL[O][i][\textlambda][#1]}
\newcommand{\OIall}{\forbiddenEL[O][i][\textlambda][\textlambda][6300,][6364]}

\newcommand{\HeI}{\permittedEL[He][i]}
\newcommandx{\HeIL}[1][1=5875]{\permittedEL[He][i][\textlambda][#1]}

\newcommand{\OIperm}{\permittedEL[O][i]}
\newcommand{\OIres}{\permittedEL[O][i]}
\newcommandx{\OIresL}[1][1=8446]{\permittedEL[O][i][\textlambda][#1]}

\newcommand{\NII}{\forbiddenEL[N][ii]}
\newcommandx{\NIIL}[1][1=6583]{\forbiddenEL[N][ii][\textlambda][#1]}
\newcommand{\NIIall}{\forbiddenEL[N][ii][\textlambda][\textlambda][6548,][6583]}

\newcommand{\SII}{\forbiddenEL[S][ii]}
\newcommandx{\SIIL}[1][1=6716]{\forbiddenEL[S][ii][\textlambda][#1]}
\newcommandx{\SIIall}[2][1=6716,2=6731]{\forbiddenEL[S][ii][\textlambda][\textlambda][{#1},][#2]}

\newcommand{\SIII}{\forbiddenEL[S][iii]}
\newcommandx{\SIIIL}[1][1=9532]{\forbiddenEL[S][iii][\textlambda][#1]}
\newcommandx{\SIIIall}[2][1=9069,2=9532]{\forbiddenEL[S][iii][\textlambda][\textlambda][{#1},][#2]}

\newcommandx{\OIIAuL}[1][1=7325]{\forbiddenEL[O][ii][\textlambda][#1]}
\newcommand{\OIIAuall}{\forbiddenEL[O][ii][\textlambda][\textlambda][7319--][7331]}

\newcommandx{\CIIFIRL}{\forbiddenEL[C][ii][\textlambda][158\,\mum]}

\newcommand{\NaI}{\permittedEL[Na][i]}
\newcommandx{\NaIL}[1][1=5890]{\permittedEL[Na][i][\textlambda][#1]}
\newcommand{\NaIall}{\permittedEL[Na][i][\textlambda][\textlambda][5890,][5896]}

\newcommand{\CaII}{\permittedEL[Ca][ii]}
\newcommandx{\CaIIL}[1][1=3934]{\permittedEL[Ca][ii][\textlambda][#1]}
\newcommand{\CaIIall}{\permittedEL[Ca][ii][\textlambda][\textlambda][3934,][3968]}

\newcommandx{\FeIIperm}{\permittedEL[Fe][ii]}
\newcommandx{\FeII}{\forbiddenEL[Fe][ii]}
\newcommandx{\FeIIL}[1][1=5159]{\forbiddenEL[Fe][ii][\textlambda][#1]}
\newcommandx{\FeIIall}{\forbiddenEL[Fe][ii][\textlambda][\textlambda][4359,][4414]}

\newcommand{\hda}{\ensuremath{\mathrm{H\text{\textdelta}_A}}\xspace}
\newcommand{\hga}{\ensuremath{\mathrm{H\text{\textgamma}_A}}\xspace}

\newcommand{\darkhorse}{Dark\,Horse\xspace}
\newcommand{\method}{dense-shutter\xspace}

\title[JADES \darkhorse]{JADES Dark Horse: demonstrating high-multiplex observations with JWST/NIRSpec \method spectroscopy in the JADES Origins Field}

\author[\sendemail{francesco.deugenio@gmail.com}{Questions about the spectral overlap paper (dark horse?)}{Hi Francesco!\%0A\%0Ahow are you doing? I have a question about this paper, do you mind? (Actually, two questions. The first is: why `dark horse'?).\%0ANow back to the other question, ...\%0A\%0ARegards,\%0A}{F. D'Eugenio}~et al.]{\parbox{\textwidth}{
\orcidsymb{Francesco D'Eugenio}{0000-0003-2388-8172}$^{\hyperlink{aff1}{1},\hyperlink{aff2}{2}}$\thanks{E-mail: francesco.deugenio@gmail.com},
\orcidsymb{Erica J. Nelson}{0000-0002-7524-374X}$^{\hyperlink{aff3}{3}}$,
\orcidsymb{Daniel J.~Eisenstein}{0000-0002-2929-3121}$^{\hyperlink{aff4}{4}}$,
\orcidsymb{Roberto Maiolino}{0000-0002-4985-3819}$^{\hyperlink{aff1}{1},\hyperlink{aff2}{2},\hyperlink{aff5}{5}}$,
\orcidsymb{Stefano Carniani}{0000-0002-6719-380X}$^{\hyperlink{aff6}{6}}$,
\orcidsymb{Jan Scholtz}{0000-0001-6010-6809}$^{\hyperlink{aff1}{1},\hyperlink{aff2}{2}}$,
\orcidsymb{Mirko Curti}{0000-0002-2678-2560}$^{\hyperlink{aff7}{7}}$,
\orcidsymb{Christopher N.~A. Willmer}{0000-0001-9262-9997}$^{\hyperlink{aff8}{8}}$,
\orcidsymb{Andrew J. Bunker}{0000-0002-8651-9879}$^{\hyperlink{aff9}{9}}$,
\orcidsymb{Jakob M. Helton}{0000-0003-4337-6211}$^{\hyperlink{aff10}{10}}$,
\orcidsymb{Ignas Juod{\v z}balis}{0009-0003-7423-8660}$^{\hyperlink{aff1}{1},\hyperlink{aff2}{2}}$,
\orcidsymb{Fengwu Sun}{0000-0002-4622-6617}$^{\hyperlink{aff4}{4}}$,
\orcidsymb{Sandro Tacchella}{0000-0002-8224-4505}$^{\hyperlink{aff1}{1},\hyperlink{aff2}{2}}$,
\orcidsymb{Santiago Arribas}{0000-0001-7997-1640}$^{\hyperlink{aff11}{11}}$,
\orcidsymb{Alex J. Cameron}{0000-0002-0450-7306}$^{\hyperlink{aff9}{9}}$,
\orcidsymb{St\'ephane Charlot}{}$^{\hyperlink{aff12}{12}}$,
\orcidsymb{Emma Curtis-Lake}{0000-0002-9551-0534}$^{\hyperlink{aff13}{13}}$,
\orcidsymb{Kevin Hainline}{ 0000-0003-4565-8239}$^{\hyperlink{aff8}{8}}$,
\orcidsymb{Benjamin D.~Johnson}{0000-0002-9280-7594}$^{\hyperlink{aff4}{4}}$,
\orcidsymb{Brant Robertson}{0000-0002-4271-0364}$^{\hyperlink{aff14}{14}}$,
\orcidsymb{Christina C.\ Williams}{0000-0003-2919-7495}$^{\hyperlink{aff15}{15}}$,
\orcidsymb{Chris~Willott}{0000-0002-4201-7367}$^{\hyperlink{aff16}{16}}$
\orcidsymb{William M. Baker}{}$^{\hyperlink{aff17}{17}}$,
\orcidsymb{Jacopo Chevallard}{}$^{\hyperlink{aff9}{9}}$,
\orcidsymb{A. Lola Danhaive}{0000-0002-9708-9958}$^{\hyperlink{aff1}{1},\hyperlink{aff2}{2}}$,
\orcidsymb{Yuki Isobe}{0000-0001-7730-8634}$^{\hyperlink{aff1}{1},\hyperlink{aff2}{2}}$,
\orcidsymb{Xihan Ji}{0000-0002-1660-9502}$^{\hyperlink{aff1}{1},\hyperlink{aff2}{2}}$,
\orcidsymb{Zhiyuan Ji}{0000-0001-7673-2257}$^{\hyperlink{aff8}{8}}$,
 \orcidsymb{Gareth C.~Jones}{0000-0002-0267-9024}$^{\hyperlink{aff1}{1},\hyperlink{aff2}{2}}$,
\orcidsymb{Nimisha Kumari}{0000-0002-5320-2568}$^{\hyperlink{aff18}{18}}$,
\orcidsymb{Tobias J.~Looser}{0000-0002-3642-2446}$^{\hyperlink{aff4}{4}}$,
\orcidsymb{Jianwei Lyu}{}$^{\hyperlink{aff8}{8}}$,
\orcidsymb{Eleonora Parlanti}{0000-0002-7392-7814}$^{\hyperlink{aff6}{6}}$,
 \orcidsymb{Michele Perna}{0000-0002-0362-5941}$^{\hyperlink{aff11}{11}}$,
\orcidsymb{D\'avid Pusk\'as}{0000-0001-8630-2031}$^{\hyperlink{aff1}{1},\hyperlink{aff2}{2}}$,
\orcidsymb{Pierluigi Rinaldi}{0000-0002-5104-8245}$^{\hyperlink{aff8}{8}}$,
\orcidsymb{Charlotte Simmonds}{}$^{\hyperlink{aff1}{1},\hyperlink{aff2}{2},\hyperlink{19}{19}}$,
\orcidsymb{Yang Sun}{0000-0001-6561-9443}$^{\hyperlink{aff8}{8}}$,
\orcidsymb{Hannah \"Ubler}{0000-0003-4891-0794}$^{\hyperlink{aff20}{20}}$,
\orcidsymb{Giacomo Venturi}{0000-0001-8349-3055}$^{\hyperlink{aff6}{6}}$,
\orcidsymb{Joris Witstok}{0000-0002-7595-121X}$^{\hyperlink{aff21}{21},\hyperlink{aff22}{22}}$,
\orcidsymb{Zihao Wu}{0000-0002-8876-5248}$^{\hyperlink{aff4}{4}}$
and 
\orcidsymb{Yongda Zhu}{0000-0003-3307-7525}$^{\hyperlink{aff8}{8}}$
}\vspace{0.4cm}
\\
\parbox{\textwidth}{
\hypertarget{aff1}{$^{1}$}Kavli Institute for Cosmology, University of Cambridge, Madingley Road, Cambridge, CB3 0HA, United Kingdom\\
\hypertarget{aff2}{$^{2}$}Cavendish Laboratory - Astrophysics Group, University of Cambridge, 19 JJ Thomson Avenue, Cambridge, CB3 0HE, United Kingdom\\
\hypertarget{aff3}{$^{3}$}Department for Astrophysical and Planetary Science, University of Colorado, Boulder, CO 80309, USA\\
\hypertarget{aff4}{$^4$}Center for Astrophysics $|$ Harvard \& Smithsonian, 60 Garden St., Cambridge MA 02138 USA \\
\hypertarget{aff9}{$^{5}$}Department of Physics and Astronomy, University College London, Gower Street, London WC1E 6BT, UK\\
\hypertarget{aff5}{$^{6}$}Scuola Normale Superiore, Piazza dei Cavalieri 7, I-56126 Pisa, Italy\\
\hypertarget{aff6}{$^{7}$}European Southern Observatory, Karl-Schwarzschild-Strasse 2, 85748 Garching, Germany \\
\hypertarget{aff7}{$^{8}$}Steward Observatory, University of Arizona, 933 N. Cherry Ave., Tucson, AZ, 85721, USA\\
\hypertarget{aff9}{$^{9}$}Department of Physics, University of Oxford, Denys Wilkinson Building, Keble Road, Oxford OX1 3RH, UK\\
\hypertarget{aff10}{$^{10}$}Department of Astronomy \& Astrophysics, The Pennsylvania State University, University Park, PA 16802, USA\\
\textit{\normalsize Remaining affiliations are listed at the end of the paper}
}
}

\date{Accepted 2 Apr. 2026. Received 8 Mar. 2026; in original form 17 Oct. 2025}

\pubyear{\the\year{}}

\begin{document}
\label{firstpage}
\pagerange{\pageref{firstpage}--\pageref{lastpage}}
\maketitle

\begin{abstract}
We present \jwst/NIRSpec dense-shutter spectroscopy (DSS). This novel observing strategy with the NIRSpec/MSA deliberately permits a high number of controlled spectral overlaps to reach extreme multiplex while retaining the low background of slit spectroscopy. In a single configuration over the JADES Origins Field, we opened shutters on all faint ($m_\mathrm{F444W}<30$~mag) $z_{\rm phot}>3$ candidates, prioritising emission-line science and rejecting only bright continuum sources. Using 33.6 and 35.8 ks on-source in G235M and G395M, we observed a single mask with $\sim$850 sources, obtaining spectroscopic redshifts for $\sim$540 galaxies over $2.5\lesssim z \lesssim 8.9$. The per-configuration target density in DSS mode is 4--5$\times$ higher than standard no- and low-overlap MSA strategies ($<$200 sources), with no loss in redshift precision or accuracy. Line-flux sensitivities are 30 percent lower at fixed exposure time, matching the expected increase in background noise, but the gain in survey speed is 5$\times$ in our setup, more than justifying the penalty. The measured line sensitivity exceeds NIRCam/WFSS by at least $\sim5\times$ ($\sim25\times$ in exposure time) at $\lambda\sim4~\mum$, demonstrating that DSS is a compelling method to gain deep, wide-band spectra for large samples. Crucially, NIRSpec/MSA could deliver even higher target allocation densities than those used here. We derive \Halpha-based SFRs, gas-phase metallicities (including a large sample suitable for strong-line calibrations), and identify rare mini-quenched galaxies and broad-line AGN. DSS is immediately applicable wherever deep imaging enables robust pre-selection and astrometry, providing an efficient method to obtain large samples of faint emission-line galaxies, a compelling middle ground between the completeness of slitless surveys and the sensitivity and bandwidth of NIRSpec/MSA.
\end{abstract}




\section{Introduction}\label{s.intro}

With its unprecedented ability to obtain sensitive near-infrared spectroscopy 
from 1--5~$\mu$m, \jwst \citep{gardner+2023} is providing revolutionary capabilities to trace 
galaxy evolution to the earliest cosmic times. Spectroscopic redshifts and 
emission-line fluxes allow us to map the physical conditions within galaxies 
as they assemble and evolve, providing insights into gas-phase metallicities, ionization 
states, and instantaneous star formation rates. Further, these observations 
enable us to dissect the role of active galactic nuclei (AGN) in regulating 
early galaxy growth and to understand how supermassive black holes themselves 
grew during the first billion years after the Big Bang. As such, large spectroscopic surveys of faint early galaxies provide key insights on the early stages of galaxy growth.

\jwst offers two primary approaches for near-infrared (NIR)
multi-object spectroscopy: NIRISS or NIRCam slitless grism modes \citep{willott+2022,greene+2017} and the 
NIRSpec multi-shutter assembly \citep[MSA;][]{ferruit+2022}.
However, these approaches present a trade-off that limits survey efficiency. MSA spectroscopy achieves substantially higher sensitivity by masking away most of the background and benefits from a wider instantaneous bandwidth and larger pixels that reduce detector noise. Yet it suffers from low targeting completeness: the fixed grid of MSA shutters vignette approximately half of all sources just by chance, and the standard practice of avoiding spectral overlaps blocks yet more.
Further, complex layers of target priorities make it difficult to identify underlying bias, let alone to reconstruct unbiased samples, with a few exceptions \citep[e.g.,][]{curtis-lake+2025,degraaff+2025b}. Conversely, slitless spectroscopy achieves higher multiplex and substantially better practical completeness---every source in the field can be 
observed simultaneously, at the expense of potential source confusion. However, with nothing blocking the background from the regions of sky without targets, the sensitivity is reduced significantly. Further,  NIRCam and NIRISS grisms cover much narrower wavelength ranges per exposure. Hence, there has been a trade-off between spectral coverage, sensitivity, and multiplexing. 

Attempting to increase the MSA allocation efficiency has led to experiments that combine
NIRSpec prism spectroscopy with overlapping grating spectroscopy \citep{eisenstein+2023a,bunker+2024,maseda+2023,degraaff+2025b}.
Because the prism spectral traces are shorter than those for the gratings, these works have relied
on the prism to disentangle overlapping sources.
Here we present a hybrid approach that we believe offers improved efficiency for programmes focused on emission lines from large samples of faint galaxies. Simply put, we drop the prism requirement,
and focusing on the NIRSpec gratings we open slits on every target object that has usable centration in its slit. Typically, this results in the overlap of 5-10 spectra, but this is of little consequence, because the dispersed background, even in the $R=1000$ gratings is still fainter than detector noise. Because emission lines 
occupy a tiny fraction of the spectral footprint compared to continuum light, 
overlap of the lines is rare. We avoid targets bright enough that their dispersed continua would exceed the detector noise and thereby degrade statistical precision on other lines. This approach 
increases targeting density by factors of 4--5 compared to standard approaches 
while preserving the sensitivity advantages of slit spectroscopy.

Given the overlap of spectra, one must successfully associate the detected lines with the correct target.
This is the same problem as with slitless spectroscopy, but now with even fewer candidates (since non-targets are masked).  There are many paths to success.  First, the detection of multiple well-separated lines immediately indicates the location of the source.  Second, the astrometric location of sources along the slit is quite diagnostic: a 3-shutter slit is 15 pixels high, so requiring an astrometric match to even a few pixels is an 80~percent rejection of false positives.  Finally, one typically has photometric redshifts that inform the plausibility of line associations.

Clearly, a requirement for this mode is to have adequate pre-imaging for target selection and astrometric alignment.  This is now available in many well-studied deep fields \citep[e.g.,][]{finkelstein+2023,eisenstein+2023a,eisenstein+2023b,bezanson+2024}. One of the commonly expressed advantages of slitless spectroscopy is the serendipitous discovery of emission lines. However, we note that if a line can be detected in a slitless grism exposure, with the full background noise plus the signal losses from the disperser, then it will be detectable in a direct image in the same filter and exposure time, even without any boost from the continuum.  Indeed, emission lines are routinely detected in deep multi-band NIRCam imaging, and such sources could be included in the target lists.

In this paper, we present the results of an on-orbit pilot observation in which we use a single MSA configuration to target all faint galaxies with photometric redshift above 3, reaching 30$^{\rm th}$ magnitude in F444W. Of the $\sim 2,500$ targets in the MSA footprint, $\sim 850$ were observed on a single configuration, and we obtained secure redshifts for 539 of these.
Unlike single-filter grism spectra, these spectra cover a wide wavelength range (1.6--5.3~\mum with two gratings), and hence are useful for analyses that require multiple emission lines.  Because the background is dispersed, the line detection limits are much fainter than with a wide-band grism, about a 5-fold decrease in error or 25-fold decrease in exposure time.

Our scientific purpose in these observations was to augment the spectroscopic coverage of the JADES Origins Field \citep[JOF;][]{eisenstein+2023b}, a deep field embedded in the broader JADES survey of GOODS-S \citep{eisenstein+2023a,rieke+2023,bunker+2024}. The JOF has exceptional NIRCam coverage, with 15 filters totalling about 350 hours of exposure time.  A partially overlapping NIRCam pointing from the OASIS programme (PIs: Looser \& D'Eugenio, program ID 5997) extends the region of medium-band coverage.  Given this imaging resource, we particularly sought to focus on galaxies above redshift 3, where the high equivalent width \Halpha and \OIIIL lines can yield spectroscopic redshifts even for very faint galaxies.  The spectroscopic measurements of redshift, line fluxes, metallicities, dynamics, AGN markers, and environment greatly complement the SED and morphological measurements coming from this deep field.

This paper is organized as follows. We  start by presenting the survey design, target
selection (Section~\ref{s.tgt}), and the data reduction (Section~\ref{s.datared}).
We then move to the redshift and flux measurements (Sections~\ref{s.zflux}
and~\ref{s.flux}), before showcasing the potential of our novel approach through
a series of science highlights (Section~\ref{s.sc}). We discuss the current
performance and future outlooks of this `\darkhorse' survey (Section~\ref{s.disc}), and we
conclude with a summary of our findings (Section~\ref{s.conc}).

Throughout this work, we assume the flat \textLambda CDM cosmology from
\citet{planck+2020}. All stellar masses assume a \citet{chabrier2003} initial mass
function (IMF). We adopt the solar metallicity $12 + \log(\mathrm{O/H}) = 8.69$
\citep{asplund+2009}.
All magnitudes are in the AB system \citep{oke+gunn1983}. All wavelengths are
vacuum wavelengths (but we use air wavelengths when reporting the name of optical
emission lines).

\section{Survey Design and Target Selection}\label{s.tgt}
\subsection{Observations}
\begin{figure}
  \includegraphics[width=\columnwidth]{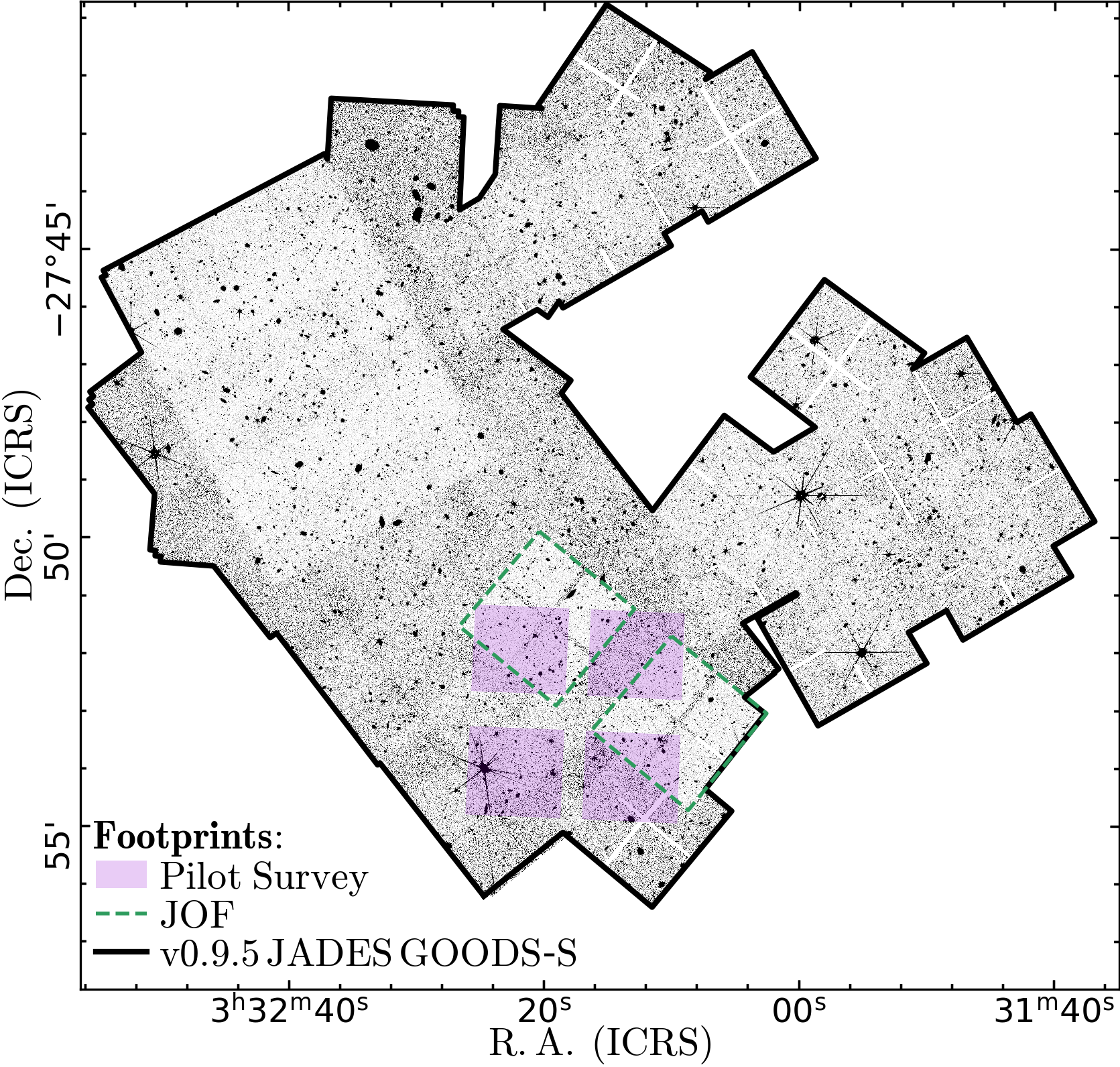}
\caption{Position of the \darkhorse pilot survey pointing (shown in pink) in relation to existing
multi-band NIRCam coverage from JADES (solid black) and to the deep
JOF field (dashed green). The NIRCam mosaics are from \citet{robertson+2026}.}\label{f.hoofprint}
\end{figure}

The Pilot Survey to showcase \method spectroscopy is \darkhorse, which
is based on
observations from \jwst PID~3215 (PIs: 
D.~J.~Eisenstein and R.~Maiolino). A fraction of PID-3215 Cycle-2 observations
failed due to MSA shorts and were re-allocated as compensation in Cycle 3.
However, when these replanned observations were lost to a telescope safing, implying a further year of delay, 
we re-designed the observations to target
the JOF and provide dense spectroscopy of the emission-line targets there. The position
of the final pointing is illustrated in Fig.~\ref{f.hoofprint}.

To focus on rest-optical lines at $z>3$, we use the G235M and G395M gratings.
These $R\sim1000$ medium resolution gratings sufficiently disperse the background and target continua so that one can overlap many spectra,
but they only marginally resolve the emission lines of normal galaxies, with a limited drop in emission-line signal-to-noise ratio (SNR) in comparison to the detector noise level.  They also provide the nominal wavelength coverage, save the gap between the two NIRSpec detectors, for any location in the MSA, and they are able to split several useful emission-line groups. These include \HeIL[1640] and \semiOIIIall,
\Hgamma and \OIIIL[4363], \Halpha and \NIIall, and \SIIL[6716] and \SIIL[6731].

The spectroscopic observations were obtained as Observation~901 on 2024 Dec. 17--18. 
They consist of a single MSA configuration with a 3-nod pattern,
repeated several times with both gratings. Nodding further reduces the \darkhorse multiplexing, by requiring one
functional shutter above and one below the allocated source; however, for \darkhorse, we
adopted this approach to help design and evaluate a master background subtraction in the future.
The two dispersers received 9.3 and 10.0 hours on source, respectively. These were
distributed in four identical sets of three nodded exposures each, with 19 groups per
integration and 2 integrations per exposure, using the \texttt{IRS}$^{\texttt{2}}$
readout \citep{rauscher+2012} and totalling 8,403 s for each of the four sets.
We note that long integrations, here 1,400 s, are preferred to suppress detector noise \citep{birkmann+2022}.
The G395M also had a single set of three nodded exposures, with 10 groups and one
integration, totalling 2,232 s, to finish using the available time allocation.

\begin{figure*}
    \centering
    \includegraphics[trim={1.cm 4.2cm 4cm 5cm},clip,width=\textwidth]{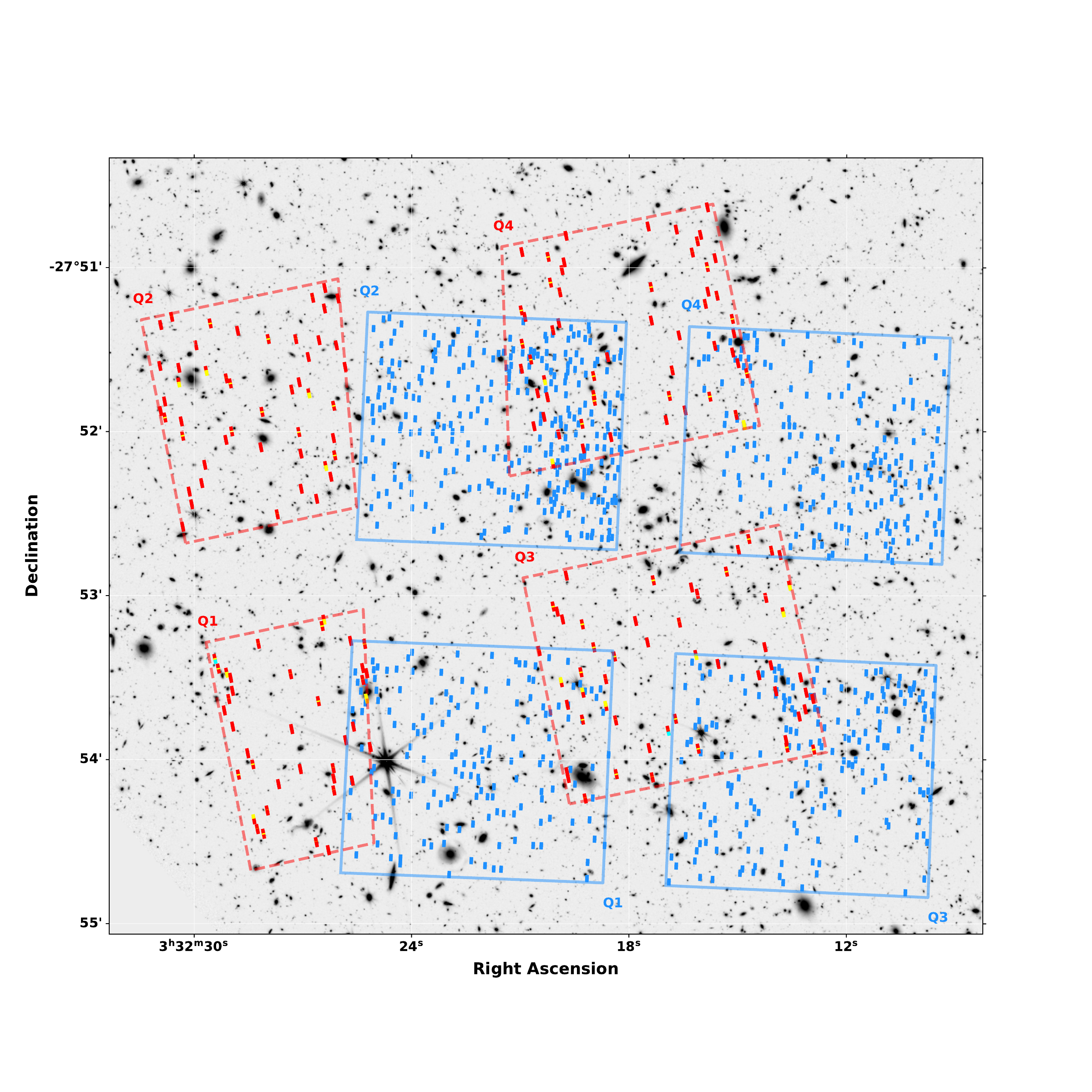}
    \caption{Comparison of a standard MSA target allocation allowing only moderate spectral overlaps
    (red slitlets; from PID~1287, \citealp{curtis-lake+2025}), and the \method method presented in this paper
    (blue). Both configurations use 3-shutter slitlets.
    The outlines of the individual MSA quadrants are shown as solid lines for \darkhorse and dashed lines
    for the standard MSA setup used in PID~1287. The shape of the latter reflects the mask design used by
    JADES to avoid truncated PRISM spectra, while \darkhorse uses the standard 
    \href{https://www.stsci.edu/files/live/sites/www/files/home/jwst/documentation/technical-documents/_documents/JWST-STScI-005921.pdf}{\texttt{NRS\_VIGNETTED\_MSA}} aperture. The background image shows a stacked F356W mosaic \citep{robertson+2026}.}\label{f.mask}
\end{figure*}

The resulting data set is very deep, with 33.6 and 35.8 ks of exposure in G235M and G395M, respectively.  
We chose to concentrate our observing time on a single deep MSA configuration rather than splitting across multiple shallower pointings to maximize sample size. This strategy leveraged the exceptional depth of the JOF imaging to probe faint galaxy populations while allowing us to test the method's limits for data reduction and deblending of faint lines.
This depth also distinguishes the new data set from the 4 JADES Medium/\jwst pointings in the same area \citep{curtis-lake+2025,scholtz+2025}, each with 9.3 ks of exposure in these dispersers.  Within JADES, only the two deep pointings, each of 25 ks per disperser, and the 3215 ultra-deep pointing with 160 ks in G395M are of comparable or deeper depth.  The Deep/\jwst pointing (program 1287) is located on the JOF; the other two are elsewhere, on the Hubble Ultra Deep Field.

\subsection{Target Selection}\label{s.tgt.ss.tgt}

The target selection in the Pilot Survey is chosen to maximise coverage at $z>3$ while being unhesitant about using the high multiplex to push to the faint galaxies available in the JOF.  From the JADES catalogue version v0.9.5 \citep{deugenio+2025a},
we require a 7-\textsigma detection in any of F277W, F335M, F356W,
F410M or F444W inside a circular aperture of radius 0.15 arcsec. To minimise detector artefacts, we also require a 7-\textsigma detection in the short-wavelengths stack, constructed by adding F090W, F115W, F150W and F200W.
This enables selection to very faint levels, as faint as 3~nJy in the JOF and 8~nJy in the flanking JADES Medium imaging (16\textsuperscript{th} percentile;
the sample medians are 13 and 27~nJy, respectively).
Although this is very faint, the anticipated line-flux limit around
$2\fluxcgs[-19]$ would itself be only a 1 nJy contribution to a NIRCam wide-band. Such a line at 3 microns in a source with a flux of 3 nJy would have an observed
equivalent width of 2000~\AA corresponding to a rest-frame equivalent width of
440~\AA in \Halpha or 330~\AA in \OIIIL; these are routine values at high redshift \citep{endsley+2023,boyett+2023}.
This calculation reinforces the point that the strong lines at high redshifts allow
\jwst slit spectroscopy to keep pace with the deep imaging.

For the redshift cut, we use \eazy photometric redshifts \citep{brammer+2008}
obtained as described in \citet{hainline+2024}. We impose $z_\mathrm{phot}>3$ and
redshift uncertainty $\sigma_z/(1+z)<0.1$, where $z_\mathrm{phot}$ is $\texttt{z\_a}$
and $\sigma_z$ is \texttt{u68}$-$\texttt{l68} from the updated catalogue of
\citet{hainline+2024}.
We stress that the medium-band imaging available in this
field combined with the strong lines of high-redshift galaxies tends to yield
unusually accurate photometric redshifts; of course, we can test this after the
fact, as presented in \S\ref{sec:photoz}.

In \method mode, avoiding bright sources is essential, because the wide spectral
coverage of the long-pass filters means that spectral overlaps have a long footprint
in the dispersion direction of the detector. To reject bright continuum sources, we
require that the minimum flux between F277W, F356W and F444W is $F_\nu < 300$~nJy
(AB mag 25.2).  
This use of minimum seeks to avoid the rejection of galaxies with strong emission lines that might boost 1 or 2 filters.
This removes 2.5~percent of the targets from the parent sample. Unfortunately, two
stars were also included by this selection, because they mimic F090W dropouts
(providing high-precision, inaccurate redshifts) and their red flux is already in
the Rayleigh-Jeans tail, eluding our continuum flux cut.

Most selected sources have weight 1 for MSA design. However, to ensure a guaranteed science return
in this experimental observation,
we upweight a number of promising sources, giving them weight 10. These consist
of `Little Red Dot' AGN candidates \citep[LRDs;][]{matthee+2024} and $z>7$ galaxies
from \citet{hainline+2024}.
A single notable source, JADES-GS-z13-1-LA \citep[][NIRSpec ID 13731]{witstok+2025}
was allocated manually, with weight 100. This galaxy is a remarkable \Lyalpha emitter
spectroscopically confirmed at $z=13.1$ with the \jwst/NIRSpec prism, so our
observations provide increased depth to attempt a detection of
\Lyalpha and rest-frame UV lines in the gratings \citep{witstok+2026}.
After the allocation procedure, we assigned 854 targets for observation, of which
28 targets are located in already occupied slitlets. The location of allocated slitlets
is shown in Fig.~\ref{f.mask}. For targets with weight 10, the allocated fraction
was 26/42, 1.8 times higher than unweighted targets.

\begin{figure}
  {\phantomsubcaption\label{f.zmag.a}
   \phantomsubcaption\label{f.zmag.b}
   \phantomsubcaption\label{f.zmag.c}}
  \includegraphics[width=\columnwidth]{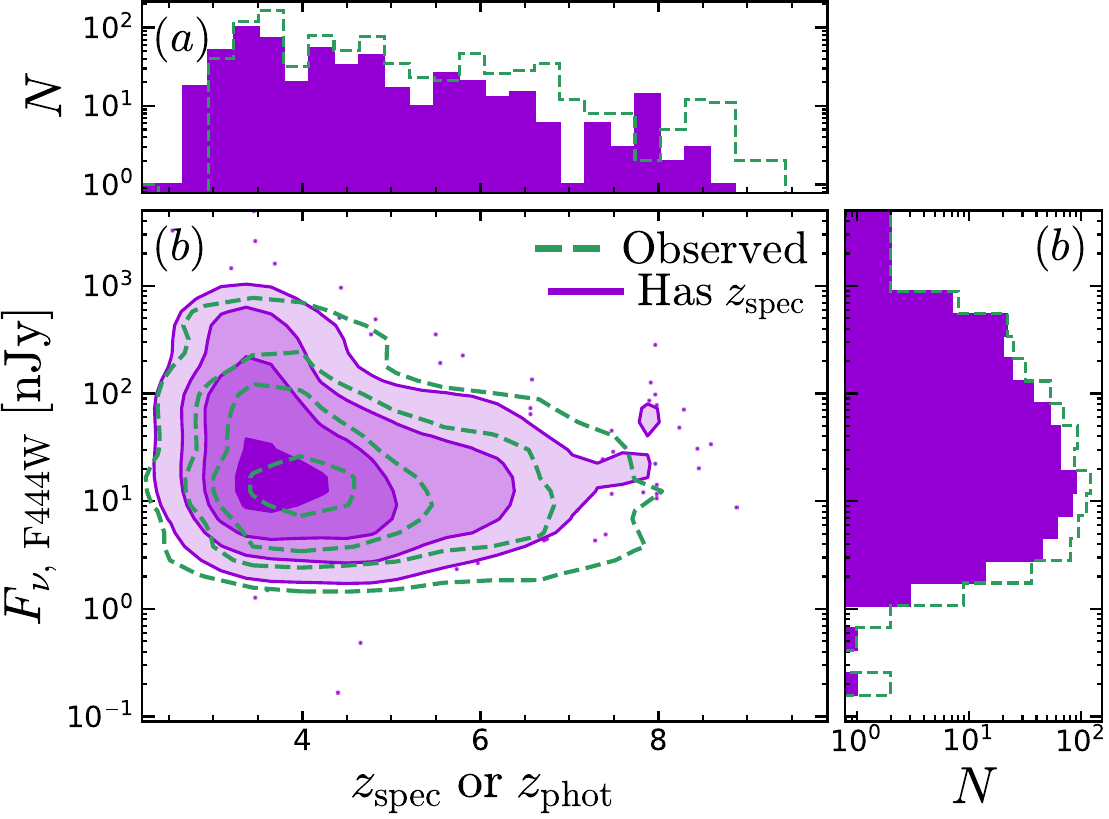}
\caption{Distribution of the Pilot Survey galaxies in the redshift--magnitude
  plane (dashed green), compared to the subset that had successful spectroscopic redshifts
  (purple). Dots are individual sources outside the outermost contour.
  We stress the high success rate to even 3 nJy continuum sources.
  }\label{f.zmag}
\end{figure}

In Fig.~\ref{f.zmag},
we show
the allocated sample 
as well as the objects for which we were able to obtain spectroscopic redshifts, reaching levels of 2--3 nJy.

In Section~\ref{s.tgt.ss.overlaps}, where we evaluate the effect of spectral overlaps, we also consider the photon noise added by overlapping source spectra. For this, we
use the typical brightness
of the allocated targets, measured inside circular apertures of 0.15-arcsec radius. The aperture
area is 20~percent smaller than the open area of a shutter, but this is
compensated by the fact that shutters (0.2-arcsec width) enclose only 80~percent
of the circle's area. To estimate the continuum, we use the minimum (considering
both detected fluxes and 3-\textsigma upper limits) among F150W, F200W, F277W,
F335M, F356W, F410M, and F444W. This yields flux densities of 14~nJy (mean) and
5~nJy (median). Using the mean flux density in each filter, we would get higher
values ranging between 25--40 nJy in the filters above. This estimate is 2--3
times higher than the previous; the difference is a combination of spectral
slopes and emission-line contribution.

\subsection{Allocation efficiency and completeness}\label{s.tgt.ss.completeness}

The completeness of the single \darkhorse configuration is $f_C=0.34$: we allocate 854
sources out of 2,516 targets within the four-quadrant footprint of the
MSA (i.e., ignoring the 23-arcsec Q1--Q3 gap and the 37-arcsec Q3--Q4 gap).
Such completeness is remarkable, given that the maximum completeness reached in the JADES Medium/JWST survey -- for much brighter targets with lower number density -- is $f_C\sim0.45$, which
required multiple overlapping pointings \citep{curtis-lake+2025}.
Three effects set $f_C$: vignetting by the MSA bars ($f_V=0.64$),
inoperable-shutters fraction coupled with the three-shutter slitlet requirement ($f_F=0.60$), plus a user-defined exclusion zone around allocated
targets (4 shutters in the dispersion direction; $f_X\approx 0.80\text{--}0.90$). So $f_C \simeq f_V \times f_F \times f_X$. We now proceed to
estimate each term.

We evaluate these three terms at the actual pointing, because the JOF is very deep,
with over twice the target density of the surrounding JADES field (after
selection, $\rho_\mathrm{JOF} \approx 350~\mathrm{arcmin}^{-2}$ vs 
$150~\mathrm{arcmin}^{-2}$). At the assigned position angle 177.57\textdegree and
at the final NIRSpec pointing $(\mathrm{R.A.,\,Dec.}) = (3^\mathrm{h} 32^\mathrm{m}
7.50^\mathrm{s},\;-27\text{\textdegree} 53^\mathrm{m} 2.71^\mathrm{s})$,
we select 2,516 targets across the four MSA quadrants, of which 1,516 are in the JOF
(4.3 arcmin$^2$; 38~percent of \darkhorse) and 1,000 in the shallower
flanking field.

To limit aperture losses, we accept sources whose centre falls within the open-shutter area,
i.e. the inner 64~percent of the $0.27\times0.53~\mathrm{arcsec}^2$ micro-shutter area (\texttt{entire\_open} setting in \textsc{mpt}; MSA Planning Tool, \citealp{karakla+2014}). This downscales the 
allocatable counts to 975 (JOF) and 643 (non-JOF).

Not all shutters are operable, and forming three-shutter slitlets further reduces availability. We calculate the average usable-shutter fraction to be $f_F = 0.60$, which yields 589 and 388 expected allocations in the JOF and flanking field, respectively, for a total allocatable source count
of 977.

While we allow spectral overlaps, \mpt still requires a minimum inter-source spacing of four shutters along the dispersion direction. This sets an effective 3$\times$9-shutter exclusion box, inside which only one slitlet can be opened. To estimate the impact of this exclusion box, we use two approaches. First, we estimate $f_X$ 
by using the exclusion-box area of 3.86 arcsec$^2$
(equivalent radius 1.1~arcsec) to identify targeting conflicts.
Of the 1,516 candidates in the JOF, 849 have no neighbour within 1.1 arcsec; the remaining 667 
have, on average, 1.6 neighbours. For simplicity, we assume that we can assign only half of these sources, 667/2. We thus estimate $849 + (667/2) = 1,183$
allocatable targets. Outside the JOF, 684 candidates are isolated; the remainder average
1.4 neighbours, for 842 effective targets. We thus estimate the allocation completeness due to the exclusion box to be $f_X = (1,183 + 842)/2,516 = 0.80$. As a sanity check, we can compare the number of sources inside the acceptance region of operable shutters (977, estimated earlier) to the actual allocated number of 854; this gives $f_X = 0.9$. The fact that this number is actually higher than the first estimate suggests that there are no other significant contributions to the efficiency of the target allocation. The discrepancy between the two estimates can be due to the simplification of assuming we observe only half of the sources with neighbours.

These factors vary with assumptions. $f_X$ depends on the exclusion-zone size
and survey depth; we measure $f_X \approx 0.85$ in the JOF and $f_X\approx0.91$
outside. The 2$\times$ higher target density in JOF more than compensates for the
drop in $f_X$, yielding more allocations per unit area.
$f_V$ reflects the accepted in-shutter area and can approach 1.0 if positional
constraints are relaxed. Finally, $f_F$ depends on operable-shutter statistics
and slitlet length. With the current operability mask, one- and five-shutter slitlets
result in $f_F = 0.74$ and $f_F=0.5$, respectively. In addition, the slitlet size also
affects the \textsc{mpt} exclusion box. For reference, single-shutter observations
(dropping nodding) would shrink the exclusion box (from $3\times9$ to 1$\times$9 shutters).
For our parent sample, this raises the calculated $f_X$ from 0.8 to $0.88\text{--}0.92$.
Together with the increased $f_F=0.74$,
this would yield $f_C\approx 0.40$, roughly 30~percent more targets per MSA configuration than
this pilot survey.

\subsection{Spectral overlaps}\label{s.tgt.ss.overlaps}

To calculate the penalty due to spectral overlaps, we compare the standard
noise terms in standard MSA spectroscopy, background and detector noise,
to the additional noise due to spectral overlaps.

Each 1,400-s integration with \texttt{IRS}$^{\texttt{2}}$ readout incurs
detector noise $\sigma_{\texttt{IRS}^{\texttt{2}}} \approx
6.3\text{--}7.4~\mathrm{e^{-}}$ \citep[][interpolated from their
table~3]{birkmann+2022}. This term also includes noise due to
dark current. This is generally the dominant noise term for
faint-object grating spectroscopy with NIRSpec/MSA.

The second important noise source is photon noise from the sky background.
In December, JOF incurs $\mu_\mathrm{B}\approx66\text{--}130$~nJy per NIRSpec
detector pixel at $\lambda=2\text{--}4.5$~\mum, appropriate for 0.1-arcsec pixels
and 0.2-arcsec wide shutters \citep{rigby+2023}.
The additional photon noise per pixel corresponds to $\sigma_{\rm F} = 2\text{--}3.7~\mathrm{e}^{-}$,
where we have used
\begin{equation}\label{eq.flux2elec}
   \sigma_\mathrm{B}^2 \equiv \mu_\mathrm{B} \; \dfrac{\mathrm{c}}{\lambda^2} \; \Delta\,\lambda \; \left(\dfrac{\mathrm{c\,h}}{\lambda}\right)^{-1} \; A \; PCE(\lambda) \; \Delta\,t,
\end{equation}
where $\Delta\,\lambda \approx 6\text{--}17.6~\AA$ is the size of a NIRSpec detector
pixel, $A$ is the telescope area, and $\Delta\,t =
1,400$~s is the integration time. The photon-to-electron conversion efficiency
PCE also includes OTE losses; we use the values based on in-flight measurements
\citep{giardino+2022}.

Finally, we estimate the photon noise from the continuum, which given our mean
flux density of 14~nJy (Section~\ref{s.tgt}), amounts to
$\sigma_{\rm S} = 1.5\text{--}2.0~\mathrm{e}^{-}$ (Eq.~\ref{eq.flux2elec}).
We assume this extra noise is spread in the spatial direction in the same manner as
the main source, and we do not account for second- and higher-order spectra, whose contribution to the
noise is lower (though in this pilot survey some rows are affected by bright targets).

Our reference noise is thus $\sigma_\mathrm{MSA}^2 = 
\sigma_{\texttt{IRS}^{\texttt{2}}}^2 + \sigma_{\rm B}^2 + \sigma_{\rm S}^2$ (see also Eq.~\ref{eq.noise} below).

In \darkhorse, 854 targets opening 2,562 shutters in the 342 rows of the
MSA, corresponds to an average of $N_\mathrm{OB} = 7.5$ overlapping background spectra per
pixel on the detector.
So overlaps cause an effective background $\mu_{\rm OB}$ which 
is 7.5 times higher than the standard background $\mu_{\rm B}$.
This is a conservative estimate: for G235M, the reddest end of the wavelength range after 3.2~\mum is significantly fainter,
due to the grating efficiency decreasing away from blaze wavelengths.
For G395M, the detector efficiency drops to zero at $\lambda \gtrsim
5.5~\mum$, hence the spectral traces are noticeably shorter than a detector, and
the effective overlap should be $2\times$ lower than for G235M. In practice, since
the outer 1/3 of both NRS1 and NRS2 is unused, the G395M traces are distributed
effectively over 2/3 of the total detector area, hence the effective overlap in G395M is
$7.5 / (2 \times 2/3) = 5.6$, where the factor 2 represents the spread over two detectors.
For the rest of the article, we ignore this difference in the interest of simplicity.

Overlapping spectra also incur additional photon noise from overlapping sources.
With 854 targets over 342 rows and two detectors, the number of source overlaps per
shutter is $N_\mathrm{OS} = 2.5 = N_\mathrm{OB}/3$,
so $\mu_{\rm OS} = 2.5 \mu_{\rm S}$, where we divide by three since each 3-shutter slitlet contains typically only one source.

Overall, the noise ratio between \method and standard MSA can be estimated
from
\begin{equation}\label{eq.noise}
\begin{split}
  \dfrac{\sigma_{\rm DH}^2}{\sigma_{\rm MSA}^2} & = \dfrac{
        \sigma_{\texttt{IRS}^{\texttt{2}}}^2 + N_\mathrm{OB}\,\sigma_{\rm B}^2 + N_\mathrm{OS}\,\sigma_{\rm S}^2}{
        \sigma_{\texttt{IRS}^{\texttt{2}}}^2 + \sigma_{\rm B}^2 + \sigma_{\rm S}^2}\\
        & = 1 + \dfrac{(N_\mathrm{OB}-1)\,\sigma_{\rm B}^2 + (N_\mathrm{OS}-1)\,\sigma_{\rm S}^2}{\sigma_{\rm MSA}^2}.
\end{split}
\end{equation}
By comparing $\sigma_{\texttt{IRS}^{\texttt{2}}}$, $\sigma_\mathrm{B}$ and
$\sigma_\mathrm{S}$, we conclude that in \method spectroscopy the dominant noise
source is still detector noise, just like in non-overlapping NIRSpec spectroscopy.
This outcome depends on our configuration, sample properties, and background.
With our mean target brightness, it is not until an average overlap of
10 (30~percent larger than our allocation) that the overlap noise reaches a level comparable to the
detector noise.
Conversely, holding constant the degree of overlaps,
one needs a mean sample brightness of 120--280~nJy in F277W--F444W before the
overlap noise due to the sources reaches the detector noise. Finally, while
background noise is generally sub-dominant in NIRSpec grating spectroscopy, this
can change when combining many spectral overlaps with higher background, e.g. in fields close to the Ecliptic.

With our same-pixel background subtraction strategy, based on three nods,
one must further add the photon and detector noise due to the subtracted background,
which are only $\sqrt{2}$ lower than for the pixel under consideration.
This further reduces the gap between \method
spectroscopy and non-overlapping allocations.

\begin{figure}
  \includegraphics[width=\columnwidth]{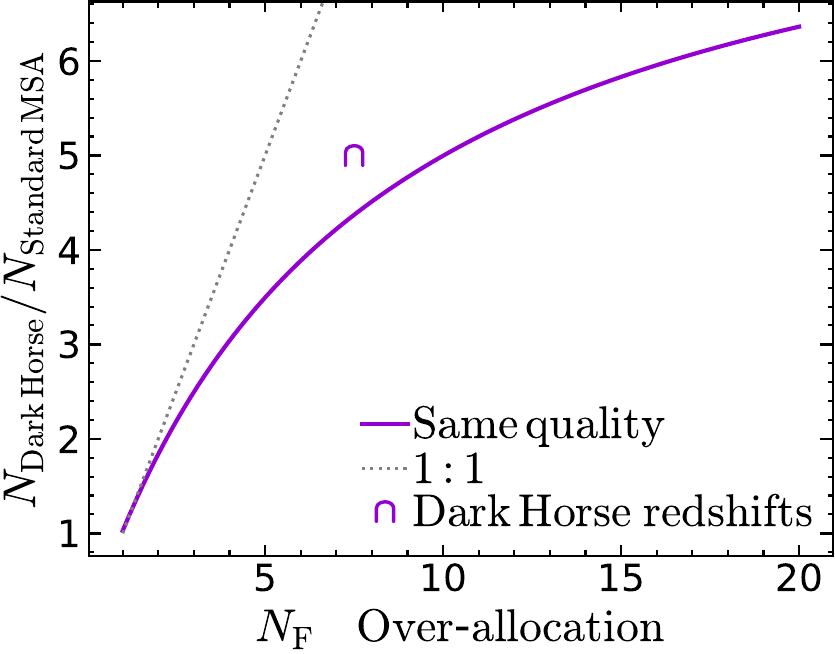}
\caption{Number of sources in \method mode
relative to standard MSA allocation, requiring the same sensitivity, and correcting
\method for the exposure time penalty due
to spectral overlaps. Since overlaps incur higher photon noise (in our case,
primarily from the background), achieving the same data quality requires longer
integration, increasing with the square-root of the number of allocated sources.
In practice, confusion will eventually cause a maximum in the curve. The single
datapoint is the empirical estimate from \darkhorse, based solely on the number
of successful redshifts (Section~\ref{s.zflux}).}\label{f.speed}
\end{figure}

With our mask design -- and ignoring sub-dominant noise terms -- the increased noise
in \method spectroscopy corresponds to a single-source time penalty factor
of $(\sigma_\mathrm{DH}/\sigma_\mathrm{MSA})^2 \approx 1.7$
relative to non-overlapping MSA spectroscopy. This estimate is validated
empirically in the next section. At the blue and red ends of the wavelength
range, the time penalty increases: 2.5 at $\lambda=1.5~\mum$ and 2.8 at
$\lambda=5.3~\mum$; this is due to the increasing sky background, which means these wavelengths
approach the background-dominated regime with fewer spectral overlaps than 2--4~\mum.
In any case, accepting this penalty is extremely advantageous, due to the large increase in
sample size, resulting in remarkably faster survey speed (Fig.~\ref{f.speed}
and Section~\ref{s.zflux}). Fig.~\ref{f.speed} shows the factor increase in sample size as a function of how many overlaps one allows. This calculation takes account of the increase in noise as a result of overallocation, thus giving an indication of the gain in efficiency allowed by \method.

\section{Data Reduction}\label{s.datared}

\begin{figure}
  {\phantomsubcaption\label{f.cts.a}
   \phantomsubcaption\label{f.cts.b}}
  \includegraphics[width=\columnwidth,trim={4.cm 4cm 2.5cm 2cm},clip]{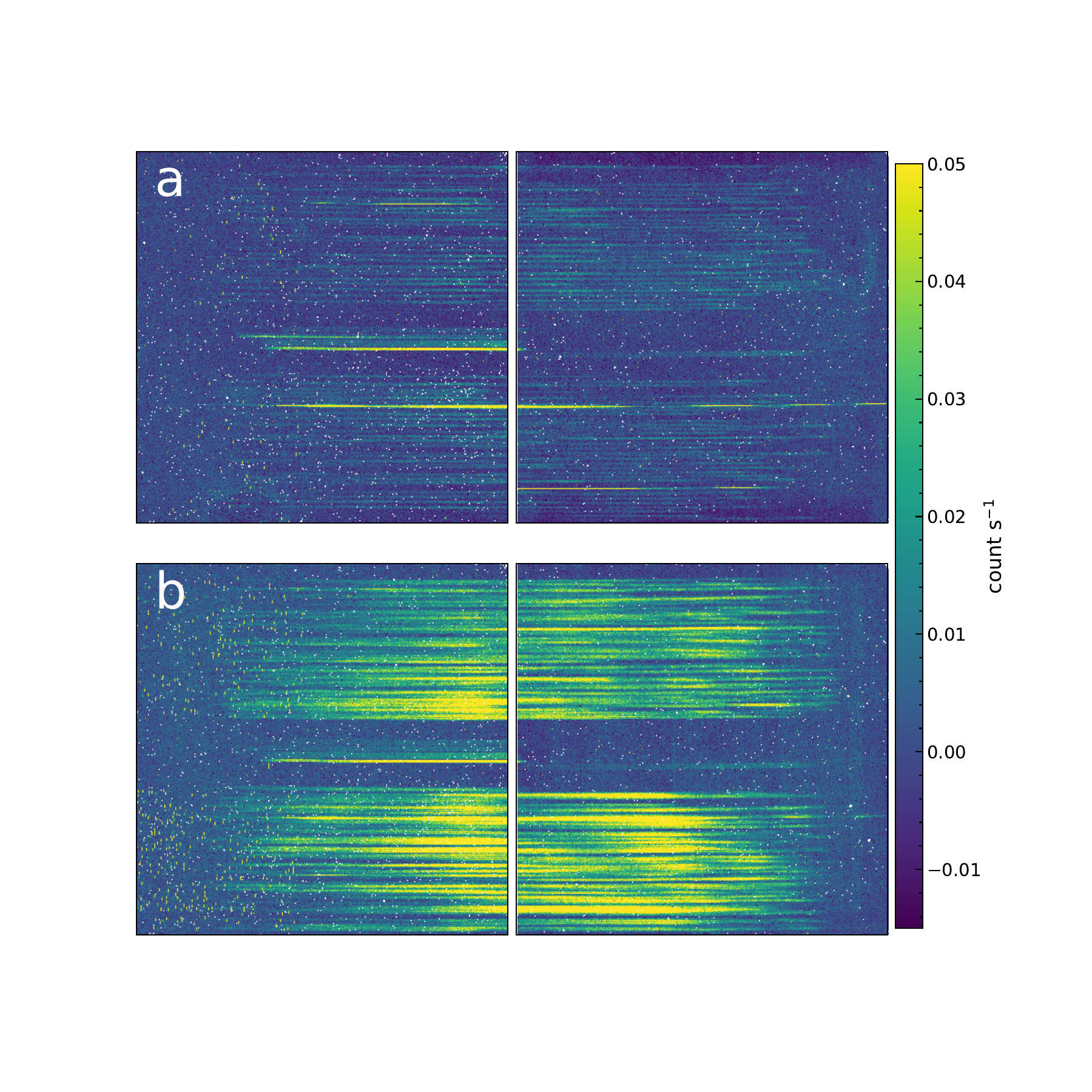}
\caption{
  Count-rate maps from PID~3215 Obs~1 (Panel~\subref{f.cts.a}) and \darkhorse (PID~3215, Obs~901; Panel~\subref{f.cts.b}).
  The two observations use the same integration per exposure (Section~\ref{s.tgt}), but each
  mask in Obs~1 allocates $\sim150$ targets, while \darkhorse allocates 854.
  The significantly higher target density of the \method comes at the cost of a somewhat brighter background.
  The bright spots on the left half of the NRS1 detector are zero-order slit images.}\label{f.cts}
\end{figure}

The data reduction procedure followed the standard methods developed by the JADES
collaboration, based on work by the ESA NIRSpec Science Operations Team. The
original algorithms have been extensively updated, as described in the JADES
spectroscopic data-release articles \citep{bunker+2024,deugenio+2025a,scholtz+2025}.
The static calibrations used are based on the context file version 1413.
The current pipeline does not automatically reduce objects falling in background shutters (28/854 sources).
We perform nod background subtraction and point-source path-loss corrections. We
refer the reader to the JADES Data Release 4 article for more information
\citep{scholtz+2025}.
Separately, we also perform a reduction with no background subtraction, to assess the
feasibility of such a strategy for emission-line science surveys (Section~\ref{s.flux}).
To illustrate the target density, we show a reduced detector
image in Fig.~\ref{f.cts}, where we compare a G235M observation from a
standard MSA configuration (panel~\subref{f.cts.a}) to the \darkhorse
observation (panel~\subref{f.cts.b}).

\begin{figure}
  \includegraphics[width=\columnwidth]{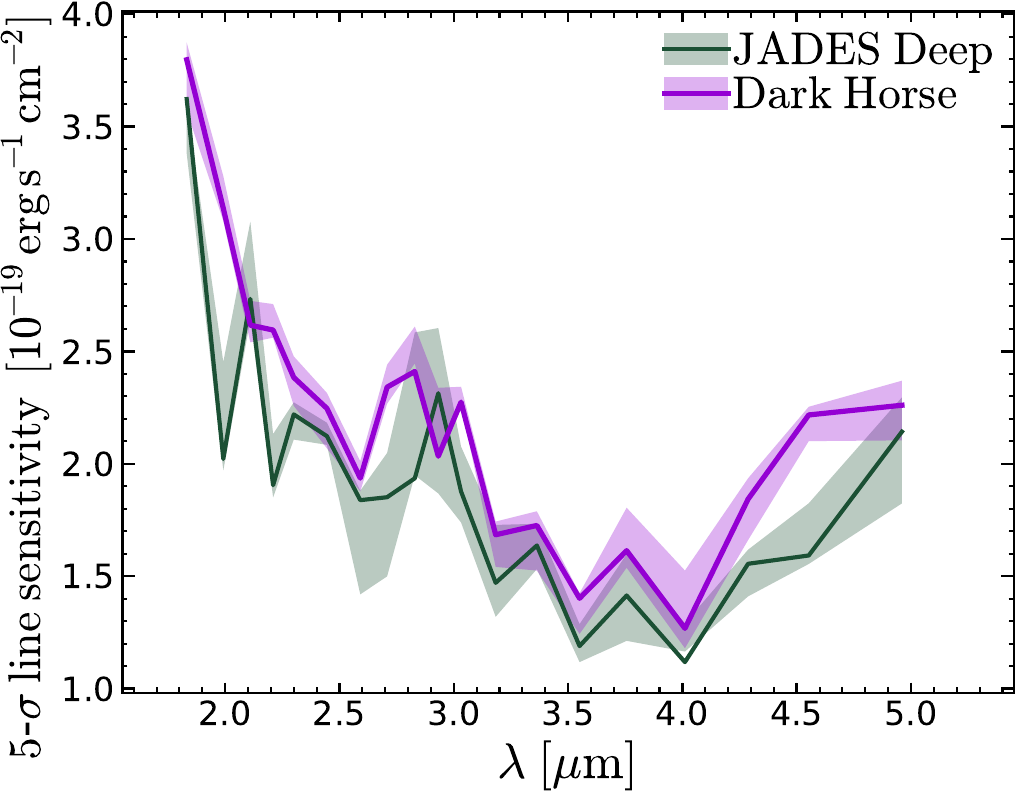}
\caption{The emission-line sensitivity of \method spectroscopy (purple) is comparable to that of
traditional NIRSpec/MSA observations of similar exposure time from JADES Deep (green). The curves
show the 5-\textsigma sensitivity in a 3-pixel boxcar extraction, estimated as the median of the noise, while the shaded regions spans the 16\textsuperscript{th}--84\textsuperscript{th} inter-percentile range.
JADES Deep used slightly shorter integration times; for an equal-time comparison
see Section~\ref{s.disc.ss.vsgrism.sss.sens}.
}\label{f.sens}
\end{figure}

The resulting 5-\textsigma emission-line sensitivity is shown in
Fig.~\ref{f.sens} (purple curve). This value is measured from the emission-line
flux and flux error of all detected emission lines ($SNR>3$), measured inside a
3-pixel boxcar extraction. We consider only lines from galaxies with redshifts.
The sensitivity is obtained by grouping the emission lines in twenty wavelength bins
with equal numbers of lines, then for each wavelength bin $i$ we fit the flux error with the formula
$\sigma_{F,i} = \sqrt{s_j + a_j \cdot F_i}$, where $F_i$ and $\sigma_{F,i}$ are the flux and flux error
of the $i\textsuperscript{th}$ emission
line in the $j\textsuperscript{th}$ wavelength bin, while $a_j$ and $s_j$ free parameters.
We then define the sensitivity as $5 \times s_j$, under the reasonable assumption that the photon-noise term $a_j \cdot F$ is sub-dominant near the sensitivity limit of our deep observations. The uncertainties are calculated for each bin
by bootstrapping the emission lines in that bin. We validate this method against sampling the
background noise in random, line-free apertures \citep[e.g.,][]{eisenstein+2023a}, and find good
agreement.
For reference, we use the same method to
calculate the sensitivity from traditional MSA observations with fewer
overlaps \citetext{green; PIDs 1210, \citealp{bunker+2024}; and~1287,
\citealp{scholtz+2025}}; these deep observations from JADES have comparable
exposure time to \darkhorse, enabling a direct performance evaluation.
The empirical noise penalty is 13~percent, calculated as the median ratio between
\darkhorse and JADES deep. This value is comparable to the estimated exposure-time penalty from 
Section~\ref{s.tgt.ss.overlaps}. There, we estimated \method to require 1.7
longer exposures to achieve the same sensitivity as standard MSA. After
accounting for the longer actual integration of \darkhorse vs PIDs 1210 and 1287
(33~ksec vs 25~ksec), we obtain an estimated noise increase of
1.14 ($\sqrt{1.7 \times 25 \div 33}$), meaning that our knowledge of the
instrument and sample enable accurate prediction of the \method performance.

\section{Redshift catalogue}\label{s.zflux}

\subsection{Redshift determination}

The redshift measurements follow the procedures outlined in \citet{deugenio+2025a}. These consist
of an initial round of visual inspection, which uses a user-movable slider to refine or change the
initial redshift guess, based on the photometric redshift from the input catalogue. This
procedure is usually accurate to within a single spectral pixel.

\begin{figure}
{\phantomsubcaption\label{f.deblend.a}
 \phantomsubcaption\label{f.deblend.b}}
      \includegraphics[width=\linewidth]{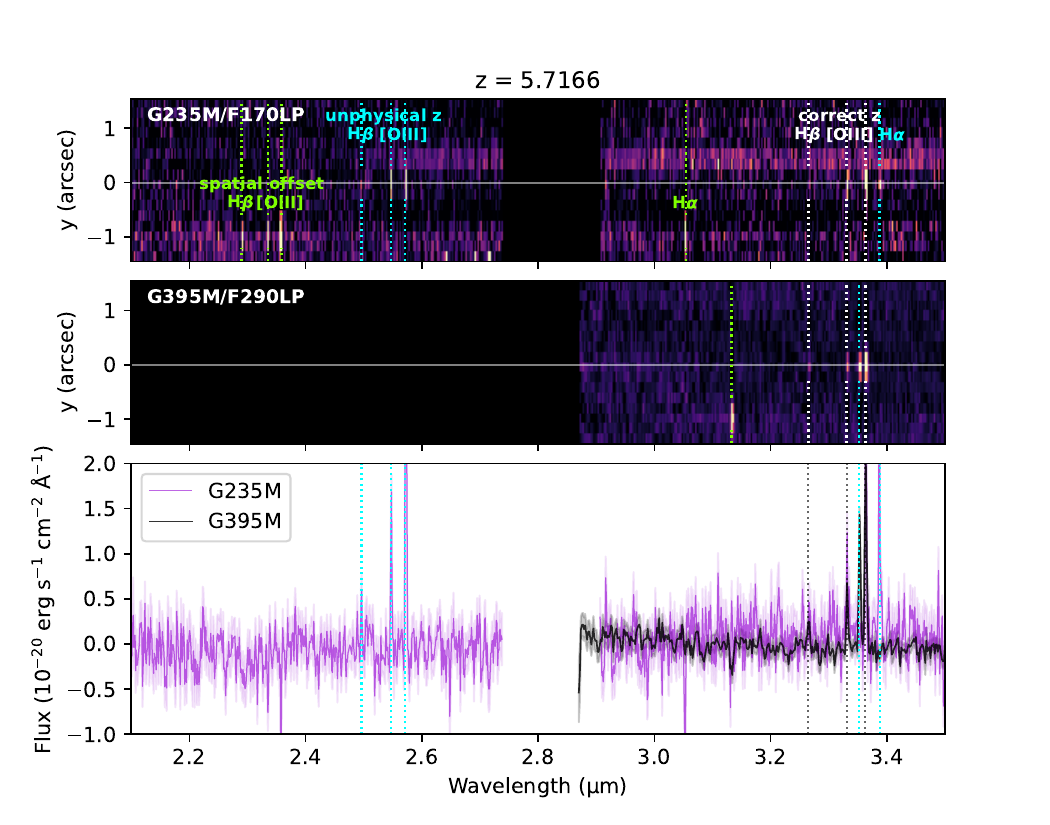}
      \caption{Example of spectral overlap identification.
      The 1-d spectrum has two readily identified \Hbeta--\OIIIall groups at
      2.6 and 3.4~\mum. In the 2-d spectrum, $y=0$ corresponds to the centre of
      the central shutter. The 2-d spectrum shows additional \Hbeta--\OIIIall emission
      near 2.3~\mum. The latter is readily excluded due to its spatial offset from
      the middle of the central shutter. The group at 2.6~\mum does not correspond
      to any redshift solution. 
      }
      \label{f.deblend}
\end{figure}

\begin{figure}
    \centering
    \includegraphics[width=\linewidth]{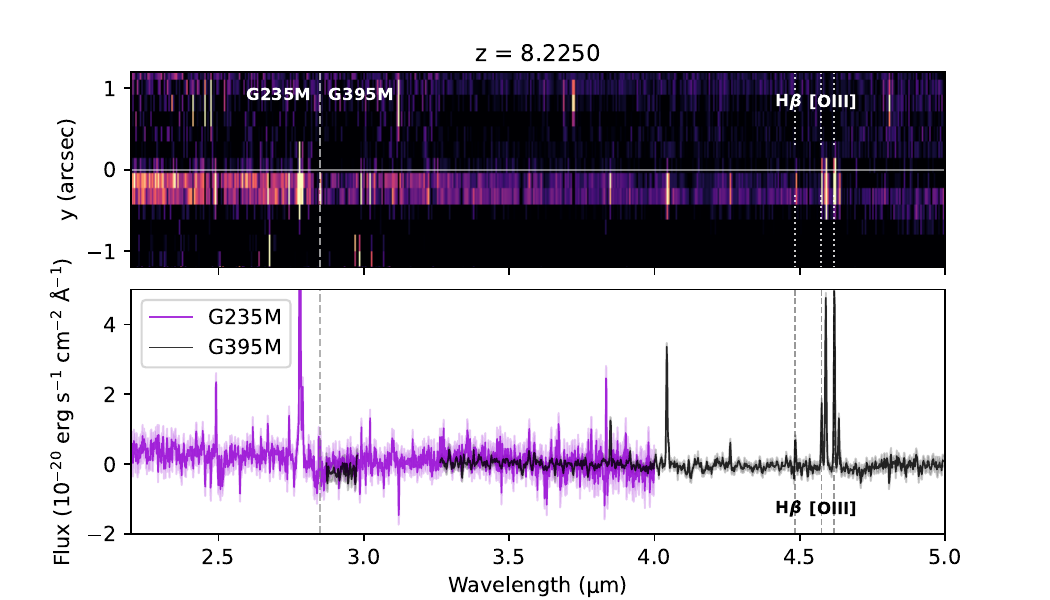}
    \includegraphics[width=\linewidth]{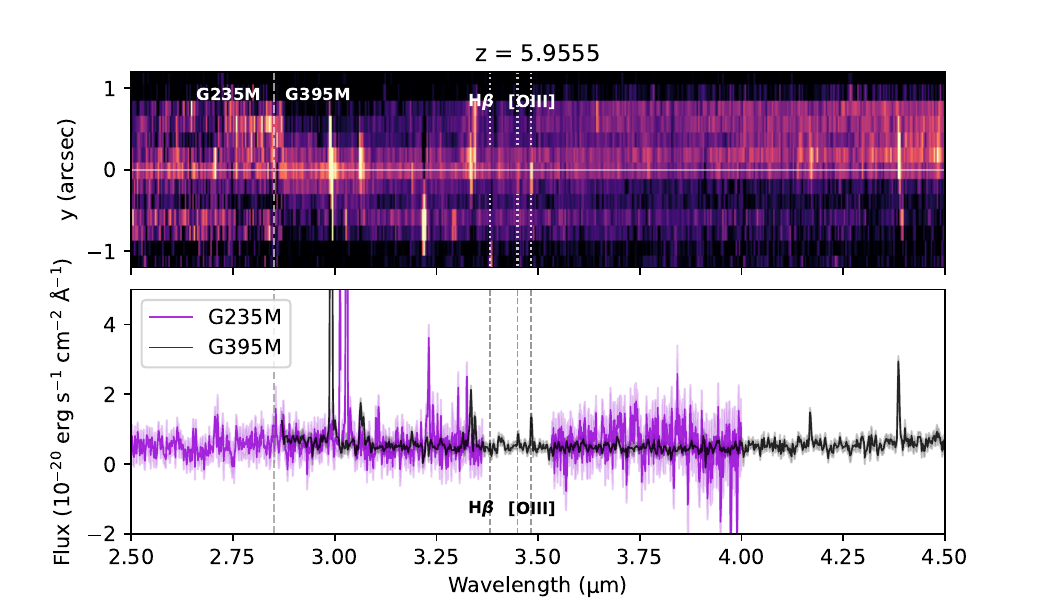}
    \includegraphics[width=\linewidth]{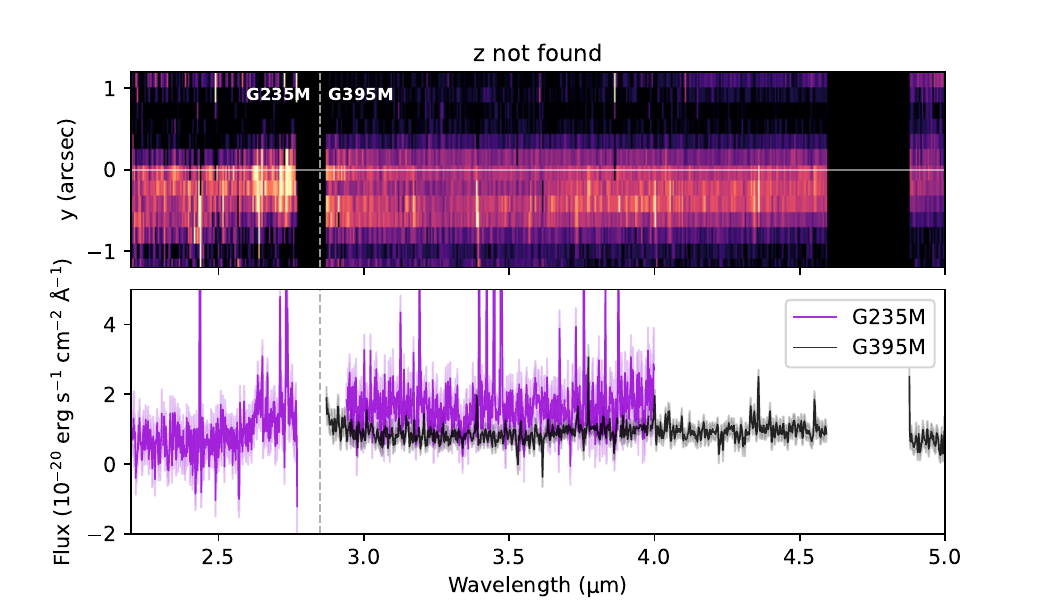}
    \caption{Three example overlaps showing a successful redshift (top), a rare example of emission-line
    overlap, i.e. a successful redshift with some corrupted line fluxes (\Hgamma--\OIIIL[4363]; middle), and an example where no redshift
    could be measured (bottom).}
    \label{f.overlap}
\end{figure}

Spectral overlaps are readily identified because they are not centred in the cross-dispersion
direction, and/or the wavelengths of multiple emission lines are not consistent with any redshift
solution (Fig.~\ref{f.deblend.a}), and/or if they are incompatible with the known SED from photometry (Fig.~\ref{f.deblend.b}).
This outcome is fully consistent with the expectations based on the small detector
footprint of emission lines. Examples of successful and unsuccessful measurements are shown in Fig.~\ref{f.overlap}

Based on the visual-inspection redshift, we use the medium-resolution line-fitting pipeline
\citep{scholtz+2025} to fit a pre-defined set of emission lines, including hydrogen and helium
recombination lines, nebular lines and the auroral line \OIIIL[4363]. From these lines, we
then define the spectroscopic redshift as follows. We establish a line priority, consisting
of \OIIIL, \Halpha, \Hbeta, \SIIIL, \HeIL[1.08\mum], \NeIIIL, \Hgamma, \Paalpha, \Pabeta. For
each target, we run through this list, and assign the redshift of the first line that has
a $SNR>5$ detection. The resulting redshift table consists 
of 519 robust redshifts from multi-line associations (and 105 tentative redshifts, identified
in visual inspection from tentative emission lines, or from strongly detected single lines).
This amounts to success rates of 63 (75) percent.

After measuring the line fluxes, we identify 535 objects with emission-line detections
above $SNR>5$, representing a success rate of 65 percent.
As a benchmark, we can use the JADES deep pointing in the HUDF (PID 1210),
which has similar gratings depth \citep{bunker+2024},
although their configuration also includes G140M and a 3 times longer exposure with
the prism.
For the comparison, we consider only galaxies with grating observations (198 of 253 targets), and
take successful redshift measurements as those with redshift flag `A', yielding a success rate of 103/198
or 52 percent \citep{bunker+2024}. Of course, being constrained by avoiding prism spectral overlaps
(and grating spectral overlaps for primary sources), the \citet{bunker+2024} result resorted to allocate
a number of sub-optimal `filler' sources. Additional losses come from the fact
that PID~1210 used three dithered pointings, resulting in a distribution of
integration times between one third, two thirds, and the full exposure time. This
complication underscores the difficulty of preserving high allocation efficiency in
a single MSA configuration -- another advantage of \darkhorse.

Overall -- comparing just the grating observations -- the \method approach delivers 25~percent higher success rate, and -- crucially -- five times more redshifts.
This is also true for the high redshift sources: PID 1210 delivered 9, 4, and 1
grating redshifts in the intervals $6\leq z < 7$, $7 \leq z < 8$, and $z\geq 8$,
respectively; the \darkhorse survey yields 34, 24 and 6.
The resulting distribution of spectroscopic redshifts is shown in Fig.~\ref{f.zmag.a},
and spans $2.5 \lesssim z \leq 8.86$.

Among the 202 redshift failures, we identify three main reasons.
\begin{compactenum}
\item Approximately one third (60/202) are invalid targets: objects with unreliable redshifts due to contaminated photometry (by neighbours or diffraction
spikes), objects that are over-shredded regions of bright foreground galaxies, and
parts diffraction spikes themselves. Of course, removing these objects from the pool of valid galaxies would increase
the success rate to 81~percent.
\item The second category of failed redshifts (63/202) are objects that are faint
($<5$~nJy), show some line excess in the photometry, but no detected emission lines at the expected location in the spectrum.
The sensitivity of these spectra may have been suppressed by the increased background or by
overlapping continuum. A better target selection without bright AGN or stars would likely recover
some of these sources.
\item The final category of redshift failures (58/202) are objects with intrinsically weak emission lines. These are identified by
visual inspection of the \eazy SED fits, where we identified clear Balmer breaks
in the photometry and little or no emission-line photometric excess.
Interestingly, this group of sources still enables compelling science
(Section~~\ref{s.sc.ss.miniq}).
\end{compactenum}
Finally, two minor classes of failed redshifts are objects where the
main emission lines are lost due to nod subtraction of bright overlapping  spectra (regardless of whether one uses all three nods or just
two nods; 4 objects) and objects where the main emission lines landed on the detector gap (17 objects).

\subsection{Comparison to photometric redshifts}
\label{sec:photoz}

\begin{figure}
  {\phantomsubcaption\label{f.zcomp.a}
   \phantomsubcaption\label{f.zcomp.b}}
  \includegraphics[width=0.95\columnwidth]{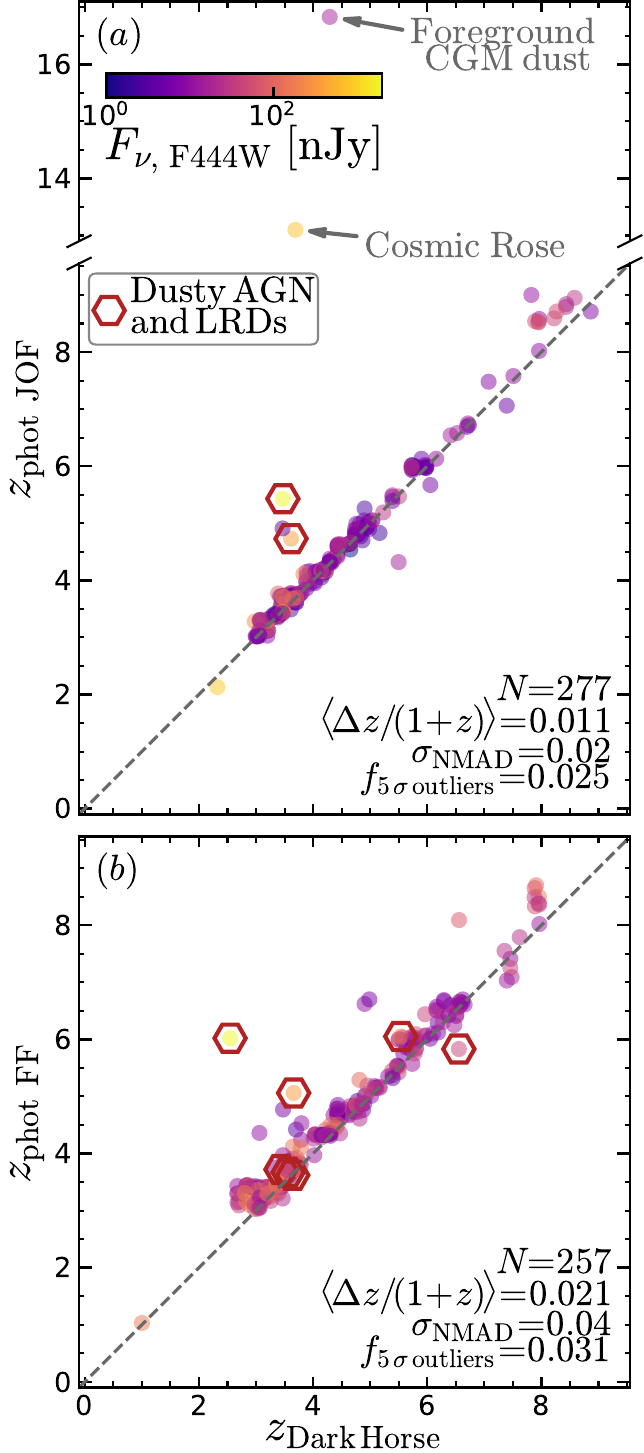}
  \caption{Comparison of the redshift accuracy in relation to photometry, splitting
  between the deeper JOF galaxies (panel~\subref{f.zcomp.a}) and the shallower flanking field
  (panel~\subref{f.zcomp.b}). Overall, the agreement is excellent, as quantified by the robust scatter measure $\sigma_\mathrm{NMAD}$ (see main text for the definition). Additionally, the few
  outliers are well understood, as explained in the text. 
  }\label{f.zcomp}
\end{figure}

\begin{figure}
  \includegraphics[width=0.95\columnwidth]{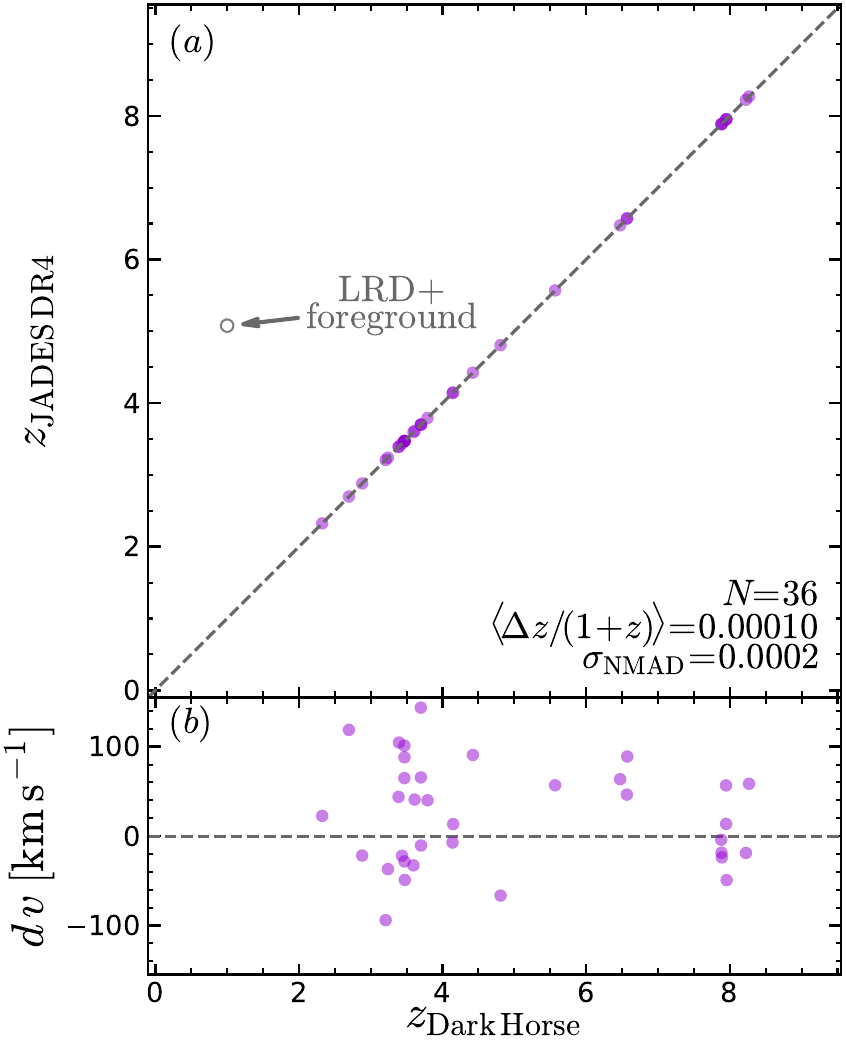}
  \caption{Comparison of the \darkhorse redshifts to known grating redshifts
  from JADES DR4. The agreement is within 10~percent of a resolution element.
  The single outlier is due to foreground contamination inside the shutter.
  }\label{f.zspec}
\end{figure}

As this is a sample of faint galaxies, with rather simple flux and photometric redshift cuts, it presents a good opportunity to test the photometric redshifts.  This is particularly important because the selection requires $z_{\rm phot}>3$.
Fig.~\ref{f.zcomp} shows the comparison of our spectroscopic redshifts to the input photometric redshifts, split between the JOF (panel~\subref{f.zcomp.a}), with deeper imaging and more complete filter set, and the flanking field (panel~\subref{f.zcomp.b}).  We find very good agreement,
with modest bias (defined as the median of $\Delta z/(1+z)$), and 3~percent 5-\textsigma outliers
-- the latter driven by the extremely small dispersion $\sigma_\mathrm{NMAD}$, calculated from the median absolute deviation 
and upscaled by the inverse of $\sqrt{2} \; \mathrm{erf}^{-1}(1/2) = 0.6745\dots$, to match the relation between median absolute deviation and standard deviation in a Gaussian distribution ($\mathrm{erf}^{-1}$ 
is the inverse error function).
The excellent agreement is a testament to the accuracy of the photometric input \citep{hainline+2024}.
The JOF displays smaller bias and scatter by a factor of 2.

Catastrophic outliers are still present in both imaging subsets. These fall into three 
main categories:
LRDs, dusty sources, and strong \OIIIL emitters near $z_\mathrm{spec}=8$.
LRDs are generally difficult to obtain good photometric redshifts for; at $z_\mathrm{spec}\approx3$, \OIIIall and \Halpha mimic
a strong Balmer break with flat optical continuum. This difficulty is not limited to faint sources,
as exemplified by GS-35453, a robust
candidate quiescent galaxy at $z_\mathrm{phot}=5.33$ \citep{alberts+2024}, which \darkhorse reveals to be an AGN at $z_\mathrm{spec}=3.66$ (Section~\ref{s.sc.ss.blagn}). Dusty galaxies
generally lack both strong spectral breaks and high-EW lines capable of anchoring a redshift; they can appear as catastrophic outliers, but this difficulty is correctly reflected
in their large confidence intervals.
One of the notable outliers in the JOF is the `Cosmic Rose' at $z=3.69$
\citep{eisenstein+2023a,alberts+2024}, the sub-mm galaxy ALESS009.1 \citep{hodge+2013,birkin+2021}.
Even after 10 hours on source, we still only detect the \Halpha--\NIIall complex, while
\Hbeta and \OIIIall remain undetected.
Finally, there is a class of strong \OIIIL emitters that suffer from a systematic bias of $
\Delta z \sim 0.5$; these objects are characterized by strong F410M--F444W excess, and weak 
detection in F090W; the latter may be driving the redshift solution to $z\gtrsim 8$, and is
perhaps connected to damped \Lyalpha absorption \citep[e.g.,][]{fujimoto+2023,asada+2025}.

Another notable outlier in Fig.~\ref{f.zcomp.a} is GS-634005, a compact, red galaxy with a
clear spectral break and a red SED.
The nominal solution at $z_\mathrm{phot} = 16.8$ would imply high dust attenuation
and large stellar mass. Instead, a clear detection of \Halpha in NIRSpec places this
galaxy at $z_\mathrm{spec} = 4.2918\pm0.0001$. The strong dust attenuation implied by the
red SED is anomalous for such a compact, low-mass system, but the proximity to the sub-mm galaxy 
ALESS010.1 at $z=3.47$ \citep[separation 3 arcsec;][]{hodge+2013} suggests we are witnessing 
foreground dust extinction in the CGM of ALESS010.1 (Section~\ref{s.sc.ss.dust}; \citealp{sun+2026}).

It is particularly remarkable that only 2 galaxies have spectroscopic redshifts below 2.5, indicating a very small rate of true low-redshift galaxies in the $z>3$ input sample.  We note that we would have detected the strong \Halpha line down to $z=1.6$, and there are plenty of rest-infrared lines (e.g., \SIIIL or \Paalpha)
that would have been in-band at lower redshift.

As a spectroscopic validation of our $z$ measurements, we consider the 43 galaxies in common between the Pilot Survey
and JADES DR4 (Fig.~\ref{f.zspec}). Of these, 40 have valid redshifts in \darkhorse, and 36 have valid redshifts in JADES DR4. The agreement is consistent within better than a
single spectral pixel, underscoring that -- as far as redshifts are concerned -- the \method
multiplexing has no cost compared to low- or no-overlap NIRSpec/MSA spectroscopy.
We note a single outlier, where we measure \Paalpha at $z_\mathrm{spec} = 1.0012$ while JADES
DR4 reports $z_\mathrm{spec} = 5.07789$. Both redshifts are accurate: this system is a LRD 
AGN at $z=5$ plus a $z=1$ interloper \citep{deugenio+2025e}; the position
of the MSA shutter in \darkhorse misses most of the LRD flux (even though faint \OIIIL
and \Halpha are still seen in the spectrum).

\section{Flux measurements}\label{s.flux}

The fluxes are measured using standard methods from the JADES Collaboration 
\citep{scholtz+2025}, consisting of a local Gaussian fit with a polynomial background. The
list of lines considered is in Table~\ref{t.lines}. Groups of spectrally adjacent emission
lines were fit simultaneously, using a common redshift and full-width at half-maximum (FWHM)
and the same continuum model; these line groups are also indicated in Table~\ref{t.lines}.

\begin{table}
\begin{center}
\caption{List of the emission lines. All wavelengths are in vacuum. Emission lines inside the same cell were fitted simultaneously using the same redshift, FWHM and continuum. }\label{t.lines}
\begin{tabular}{l|ll}
\hline
     Line(s)                   & $\lambda$ [\AA]  & Column name \\ 
\hline
     \CIVall                   & 1549.48          & \verb|C4_1549|  \\
     \HeIIL                    & 1640.00          & \verb|He2_1640| \\
     \semiOIIIall              & 1663.00          & \verb|O3_1663|  \\
\hline
     \CIIIall                  & 1907.71          & \verb|C3_1907|  \\
\hline
     \OIIall                   & 3728.49          & \verb|O2_3727|  \\
     \NeIIIL[3869]             & 3869.86          & \verb|Ne3_3869| \\
\hline
     \Hdelta                   & 4102.86          & \verb|HD_4102|  \\
\hline
     \Hgamma                   & 4341.65          & \verb|HG_4341|  \\
     \OIIIL[4363]              & 4363.44          & \verb|O3_4363|   \\
\hline
     \Hbeta                    & 4862.64          & \verb|HB_4861| \\
     \OIIIall                  & 4960.30,5008.24  & \verb|O3_5007| \\
\hline
     \HeIL[5875]               & 5877.25          & \verb|He1_5875| \\
\hline
     \OIall                    & 6302.05          & \verb|O1_6300| \\
\hline
     \Halpha                   & 6564.52          &  \verb|HA_6563| \\
     \NIIall                   & 6549.86, 6585.27 & \verb|N2_6584|   \\
     \SIIall                   & 6718.29, 6732.67 & \verb|S2_6718|, \verb|S2_6732| \\
\hline
     \HeIL[7065]               & 7067.14          & \verb|He1_7065| \\
\hline
     \SIIIall                  & 9071.10,9533.20  & \verb|S3_9069|,\verb|S3_9532| \\
\hline
     \Padelta                  & 10052.12         & \verb|PaD_10049| \\
\hline
     \HeIL[10829]              & 10832.06         & \verb|He1_10829| \\
     \Pagamma                  & 10940.98         & \verb|PaG_10938| \\
\hline
     \Pabeta                   & 12821.43         & \verb|PaB_12818| \\
\hline
     \Paalpha                  & 18755.80         & \verb|PaA_18751| \\
\hline
\end{tabular}
\end{center}
\end{table}

The depth of our observations is illustrated in Fig.~\ref{f.sens}, where we show the
5-\textsigma emission-line sensitivity as a function of wavelength.
The achieved depth is almost the same as what was obtained in the JADES deep pointings
(green line).
Overall, we demonstrate that the \darkhorse approach substantially increases the MSA
multiplex with an acceptable penalty in sensitivity. We expect these conclusions to also
apply to the higher-resolution gratings. In contrast, the prism may suffer from additional
noise, due to its lower spectral resolution and higher continuum sensitivity; the
exact performance of the prism in \darkhorse mode will be the subject of a dedicated
article.

\begin{figure}
  {\phantomsubcaption\label{f.fluxcomp.a}
   \phantomsubcaption\label{f.fluxcomp.b}
   \phantomsubcaption\label{f.fluxcomp.c}
   \phantomsubcaption\label{f.fluxcomp.d}}
  \includegraphics[width=\columnwidth]{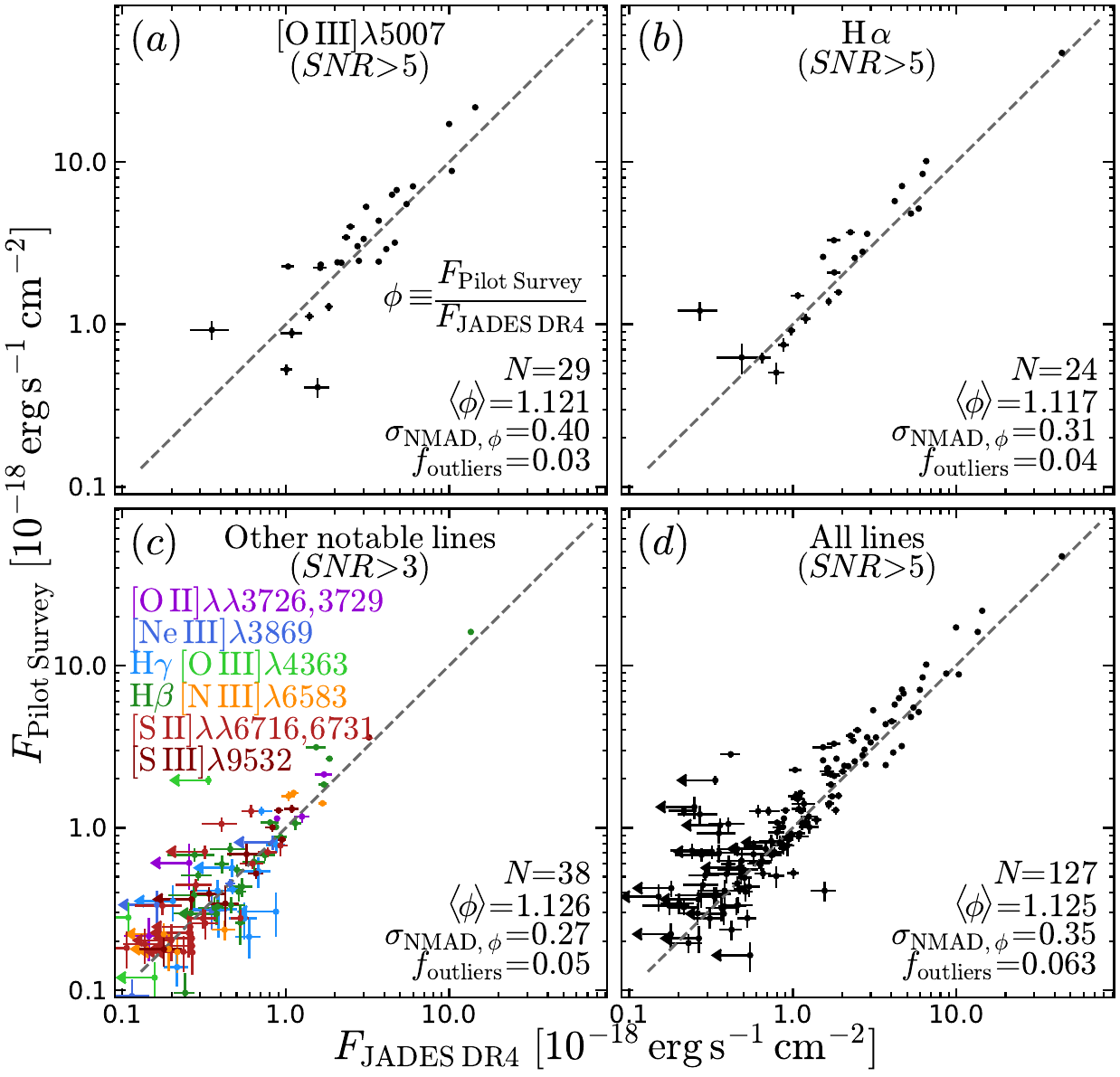}
\caption{Comparison of the flux measurements from this work to those from JADES DR4, for 43 objects in common between the two surveys. The four panels show either individual lines (\OIIIL and
\Halpha in panels~\subref{f.fluxcomp.a} and~\subref{f.fluxcomp.b}), or a selected range of 
other notable emission lines (panel~\subref{f.fluxcomp.c}), or all strongly detected lines together (panel~\subref{f.fluxcomp.d}). In each panel, we report the number of emission lines above the
specified detection threshold, the median flux ratio, the robust `NMAD' scatter about the median,
and the fraction of 3-\textsigma outliers. The results show excellent correlation; the 10~percent
flux discrepancy due to different path-loss corrections. We detect no bias between forbidden, permitted, or weak lines, which all agree with the trends observed in the full line dataset. \darkhorse spectroscopy delivers high-fidelity line fluxes.}\label{f.fluxcomp}
\end{figure}

In Fig.~\ref{f.fluxcomp} we gauge the accuracy of \darkhorse spectroscopy, using fluxes from
JADES DR4 as a benchmark. We identify 43 galaxies in common between JADES DR4 and the \darkhorse
pilot survey, which include 129 emission lines above a robust detection threshold of SNR$>5$.
The adopted SNR cut is applied to \darkhorse only, because the JADES DR4 sample includes targets
from the medium-depth tiers, which received 1-2 hours integration per grating, i.e. 5-10 times 
shallower than the Pilot Survey. This results in some line detections being only upper limits in
JADES DR4. To identify systematic differences, we split the emission lines in four subsets:
\OIIIL is representative of strong forbidden lines (Fig.~\ref{f.fluxcomp.a}); \Halpha is
representative of strong permitted lines, which can be broad in AGN (panel~\subref{f.fluxcomp.b}).
We then show a selected set of other emission lines; here we use a more generous threshold of SNR$>3$, because these lines are generally weaker than \OIIIL and \Halpha in the mass and redshift range relevant here (panel~\subref{f.fluxcomp.c}). Finally, we show all robust detections (panel~\subref{f.fluxcomp.d}).
Each panel reports the median of $\phi$, the flux ratio between the Pilot Survey flux and the JADES 
DR4 flux, with the latter adopted as ground truth.
We find 10~percent higher flux in \darkhorse
relative to JADES. This discrepancy correlates with both the intra-shutter offset
along the dispersion direction ($P=0.01$) and anti-correlates with the observed line wavelength ($P=0.03$). We attribute the discrepancy to the empirical path-loss corrections, which are reflection-symmetric along the cross-dispersion direction of the MSA, but not along the dispersion direction, thus explaining the correlation with the intra-shutter offset in the dispersion direction only \citep[see][their Appendix~B]{scholtz+2025}. The anti-correlation with wavelength is readily explained by the increasing PSF size with redder wavelengths.
Over several pointings, this bias should be effectively random, but since we are dealing with single-pointing surveys and a small overlap of 43 galaxies, the fixed pitch of the MSA introduces this systematic trend.

\begin{figure}
  \includegraphics[width=\columnwidth]{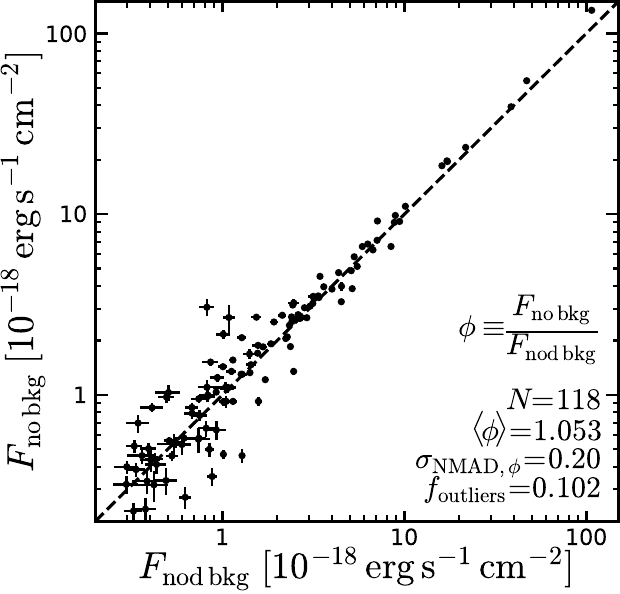}
\caption{Comparison of the default flux measurements obtained using nod background subtraction to those obtained without removing the
background, and using the emission-line fitting code to also model a local background. There
is excellent agreement. At high fluxes, the alternative reduction without background
subtraction yields higher fluxes, because it avoids self subtraction.
}\label{f.noback}
\end{figure}

As an alternative to the standard data reduction, we also analysed the emission-line fluxes
without background subtraction, leaving it to the line-fitting algorithm to remove the background.
The resulting line fluxes are shown in Fig.~\ref{f.noback}, where we find good agreement
with the default data reduction. The systematic offset to higher flux is due to avoiding
nod subtraction; this can happen in some faint sources, but is systematic for bright
sources, due to self-subtraction of spatially extended sources.
Overall, the agreement supports allocating even more microshutters, and using local methods
or a model for background subtraction.

\section{Science Highlights}\label{s.sc}

Being a method to simply increase the emission-line yield of NIRSpec/MSA, \darkhorse
naturally lends itself to most of the applications of standard NIRSpec/MSA
spectroscopy. The only notable exception is continuum science, which includes
stellar-population inference and equivalent-width measurements. Still, precise
knowledge of the redshift and emission-line fluxes enables us to supplement and improve
photometric constraints, which are available by design in \darkhorse. In the
following we present a list of science highlights, using where relevant the JADES
HUDF observations as a benchmark.

\subsection{Star-forming main sequence}\label{s.sc.ss.sfms}

The pilot survey reports the 5-\textsigma detection of 620 Balmer lines,
between \Halpha, \Hbeta, \Hgamma and \Hdelta. These can be used to estimate
accurate SFRs on short timescales of 3--10~Myr
\citep{kennicutt+evans2012}, for instance using the scalings between \Halpha
luminosity and SFR from \citet{shapley+2023}, which are appropriate for low-mass,
metal-poor galaxies. Where multiple Balmer lines are detected, we estimate directly
the dust attenuation, with the usual assumption of intrinsic Balmer ratios for
case-B recombination \citep[e.g.,][]{osterbrock+ferland2006}, electron temperature
$\Telec=10,000$~K and electron density $\nelec=500~\percm$ \citep{isobe+2023}.
Following \citet{curti+2023}, we employ the extinction curve from \citet[][hereafter:
\citetalias{gordon+2003}]{gordon+2003} to infer the dust-attenuation $A_V$. We
measure 177 $A_V$ values using the flux ratio \Halpha/\Hbeta and 51 values using
\Hgamma/\Hbeta. This is comparable to or larger than other studies of the nebular attenuation at high redshift with JWST \citep{sandles+2023,shapley+2023,woodrum+2025}.
For all other galaxies, we use the dust attenuation estimated from
spectral energy distribution (SED) models obtained with the Bayesian tool
\prospector \citep{johnson+2021}, using the public catalogue from
\citet{simmonds+2025b}.

\begin{figure}
  \includegraphics[width=\columnwidth]{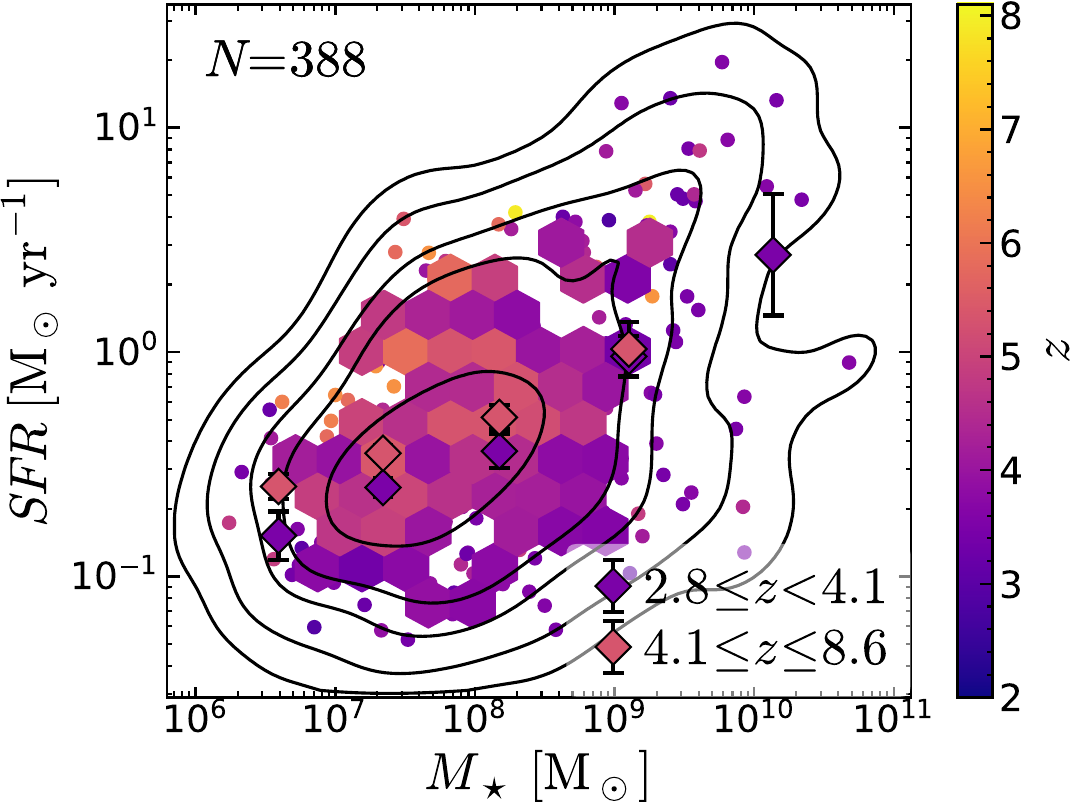}
  \caption{Redshift-collapsed star-forming main sequence from \darkhorse, at $z=2.8\text{--}8.6$. Based
  on \prospector-derived \mstar and on SFR from \darkhorse emission lines. The binned
  measurements highlight the increasing main-sequence zeropoint at earlier epochs.
  }\label{f.sfms}
\end{figure}

We illustrate the resulting star-forming main sequence (SFMS) in Fig.~\ref{f.sfms}.
Multiple \darkhorse pointings can deliver thousands of galaxies, which would enable
dissecting the SFMS in redshift, \mstar and environment -- all while using a
self-consistent selection function, high completeness, and slit-spectroscopy depth.
Last but not least, the depth of the Pilot Survey enables studying galaxies below the
SFMS, which we discuss in Section~\ref{s.sc.ss.miniq}.

\subsection{Gas-phase Metallicity}\label{s.sc.ss.metals}

\jwst has revolutionized our knowledge of emission-line galaxies at epochs earlier
than Cosmic Noon, which were only known through a handful of ground-based
observations of UV-bright or far-infrared-bright galaxies \citep[e.g.,][]{
capak+2015,inoue+,carniani+2017,stark+2015,matthee+2015}.
A number of \jwst works have uncovered the increased scatter about the metallicity
relations at $z>4$ \citep[e.g.,][]{nakajima+2023,curti+2024},
including the possible breakdown of these fundamental relations
\citep{curti+2023,curti+2024}.

We measured gas-phase metallicities adopting the strong-line calibrations for the high-z Universe presented in 
\cite{cataldi+2025}\footnote{See also \citet{sanders+2025} for recent developments.}.
The strong-line metallicities are based on the dust-corrected strong
emission-line ratios 
\begin{equation}\label{eq.strongcal}
  \begin{split}
     \mathrm{R_3}    \equiv & \dfrac{\OIIIL}{\Hbeta} \\
     \mathrm{R_2}    \equiv & \dfrac{\OIIall}{\Hbeta} \\
     \mathrm{\tilde{R}} \equiv & 0.88 \times \mathrm{log_{10}(R_3)} + 0.46 \times \mathrm{log_{10}(R_2)} \\
     \mathrm{Ne_{3}O_{2}} \equiv & \dfrac{\NeIIIL}{\OIIall} \\
  \end{split}
\end{equation}
The best-fit metallicity is found adopting a minimisation approach as described in \cite{curti+2024}, and an MCMC (with \citet{foreman-mackey+2013} implementation) is run to infer the uncertainties.  
In order to include a given diagnostic in the procedure, we request 3-\textsigma detections on the involved emission lines, 
while upper limits on \OIIall and/or \NIIL are used to discriminate among the lower- and upper-metallicity branch in case $\mathrm{R_3}$ is the only diagnostic available.
This procedure allocates 354 strong-line metallicity measurements, roughly 10  times more than PID~1210, when considering only the 36 galaxies with grating measurements; for a fair
comparison, we removed the prism-based metallicities \citep{curti+2024}, which in PID~1210
rely on 3$\times$ longer NIRSpec integration using the prism \citep{bunker+2024}.

\begin{figure}
  \includegraphics[width=\columnwidth]{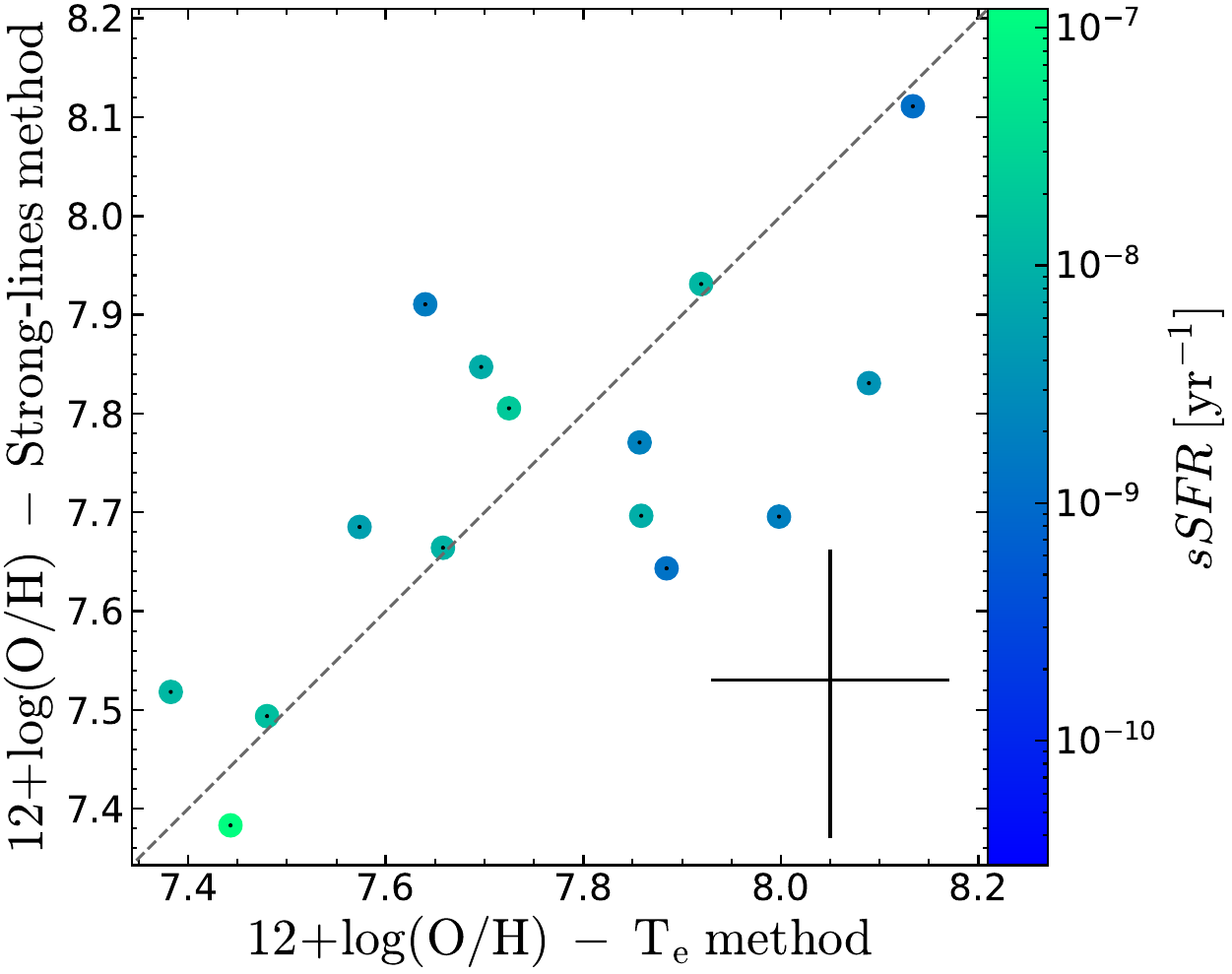}
  \caption{Comparison of gas metallicities between the direct \Telec method and the
  strong-line method, colour-coded by specific SFR.
  \darkhorse measures 18 \Telec-based metallicities (we removed four AGN with \Telec-based
  metallicities). With these numbers, one can start analysing secondary correlations,
  underscoring the advantage of \method spectroscopy in gathering large samples of faint
  emission-line measurements. The cross represents the median errorbar, including the uncertainties on \Telec.
  }\label{f.direct}
\end{figure}

In addition, we could apply the `direct' \Telec-based method \citep[e.g.,][]{
laseter+2023} for 18 galaxies with \OIIIL[4363] auroral line detection at >3-\textsigma significance, highlighting the potential of the \darkhorse approach in performing a less-biased search of this line emission in large galaxy samples (Fig.~\ref{f.direct}).
When compared to strong-lines-based metallicities for the same galaxies, \Telec-based metallicities are $\sim0.07$~dex larger on average (mean offset), with a scatter between the two measurements of $\sim0.13$~dex, consistent with the measurement uncertainties (the uncertainties on the \Telec-based metallicity take into account also the uncertainties on \Telec itself).

\begin{figure}
  \includegraphics[width=\columnwidth]{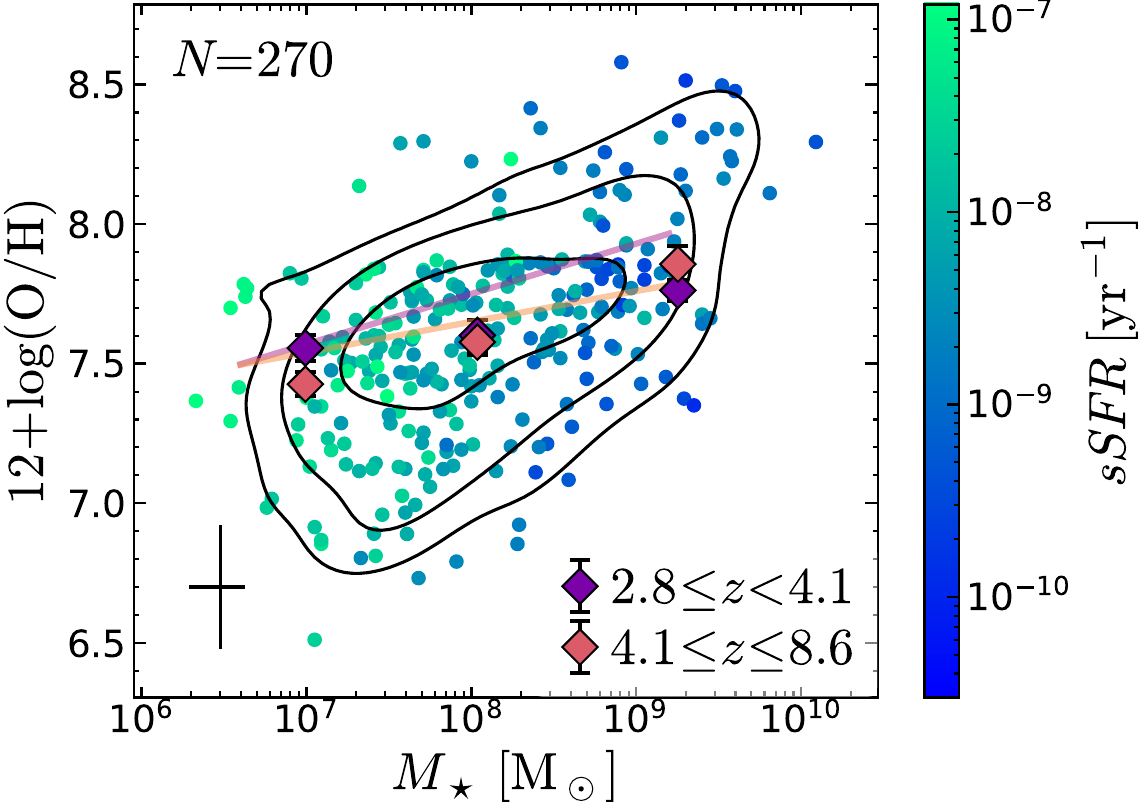}
  \caption{Mass--metallicity relation based on strong-line calibrations,
  illustrating the redshift evolution (using two equal-size subsets split by
  redshift), and the selection
  bias with star-forming activity (highlighted by the clear trend in sSFR).
  The horizontal lines are fits from \citet{curti+2024}. The offset with respect to
  our measurement reflects the different strong-line calibrations used here.
  }\label{f.mzr}
\end{figure}

With these numbers, a single pointing in \darkhorse mode delivers sufficient
galaxies to reconstruct the mass--metallicity relation at $z=3\text{--}8$
(Fig.~\ref{f.mzr}).
As for the SFMS (Section~\ref{s.sc.ss.sfms}), here too the analysis does benefit
from straightforward selection criteria. The colour coding shows the prevalence of
evolved systems (low specific SFR, sSFR) at the high-mass end (blue hues) and
starburst systems at low \mstar (green). The \darkhorse observing strategy provides
almost the same gratings depth as deep pointings, but with sufficient flexibility
to simultaneously probe $\mstar\lesssim 10^{10}~\Msun$.

\subsection{Probing CGM dust using background galaxies}\label{s.sc.ss.dust}

\begin{figure}
  {\phantomsubcaption\label{f.dust.a}
   \phantomsubcaption\label{f.dust.b}
   \phantomsubcaption\label{f.dust.c}}
  \includegraphics[width=\columnwidth]{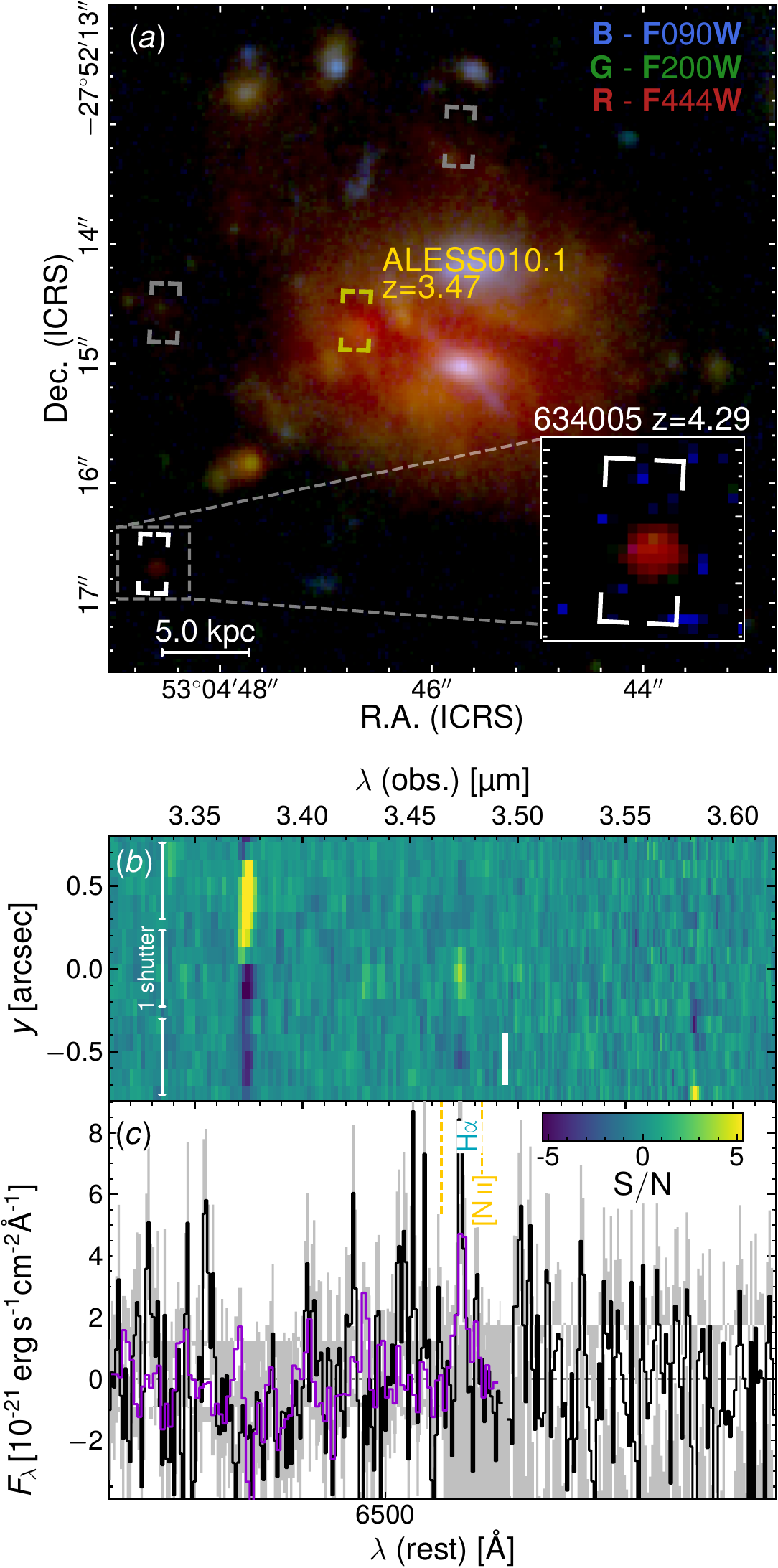}
  \caption{NIRCam false-colour image of the sub-mm galaxy ALESS010.1,
  which \darkhorse places at $z=3.47$ (panel~\subref{f.dust.a}). The bottom two panels show the grating spectra of the background galaxy 634005; the 2-d SNR map from G395M (panel~\subref{f.dust.b}) shows a clear detection in the centre of the
  shutter at $\lambda = 3.47~\mum$, also seen in the negative images (the bright
  line at $\lambda = 3.37~\mum$ is a spectral overlap from a different shutter).
  The line at 3.47~\mum is seen clearly in both the G395M and G235M 1-d spectra (black and purple lines in panel~\subref{f.dust.c}), ruling out an overlap. We identify this line with
  \Halpha at $z=4.29$ (matching \OIIIall is also seen
  in G235M, not shown here). By studying foreground dust extinction in background
  galaxies such as 634005, \darkhorse reveals the presence of dust in the CGM,
  at a distance of 20~kpc from ALESS010.1 \citep{sun+2026}. Grey shutters show \darkhorse targets without redshift.
  }\label{f.dust}
\end{figure}

The high target density of \darkhorse and the depth of our observations enable
using background galaxies to trace the CGM around massive galaxies. We report the
case study of the sub-mm galaxy ALESS010.1 at $z=3.471$ (Fig.~\ref{f.dust};
source 645014 in \darkhorse). The background galaxy 634005 is red and compact and
the photometry indicates a clear spectral break in F200W. If interpreted as a
\Lyalpha break, this would place 634005 at $z \gtrsim 16$. However, at this redshift,
the rest-frame UV slope would be unusually flat ($\beta_\mathrm{UV}>0$), and the stellar mass of the
galaxy would be unusually large. SED fitting with \prospector finds $z_{\rm phot}
= 16.4\pm0.5$ and $\log(\mstar/\Msun) = 9.3\pm0.3$ (see Section~\ref{s.sc.ss.miniq}
for our \prospector setup).
However, thanks to the depth of \darkhorse, we detect unambiguous \Halpha emission
at $z=4.29$ (Fig.~\ref{f.dust.c}), consistent with \eazy's secondary low-redshift solution,
where the spectral break is the Balmer break, made stronger by dust attenuation.

However, even this low-redshift solution seems at first unlikely: in this case, the SED
model infers a quiescent solution, with $\log(\mstar/\Msun) = 8.7\pm0.2$ and
$A_V \approx 1.4$~mag in the stellar continuum. Since dust attenuation correlates
with \mstar \citep[e.g.,][]{sandles+2023,woodrum+2025}, such a high amount of obscuration in
such a low-mass galaxy seems unlikely -- particularly for the stellar continuum.
Intriguingly, other galaxies around ALESS010.1 also display unusual levels of
attenuation, given their redshift and stellar mass. This suggests that the dust
attenuation is in the foreground, such that these low-mass galaxies are probing
dust in the CGM around the ALESS010.1, at distances of $\approx 20$~kpc.
This is smaller but comparable to the scales over which we see filamentary ionized structures
-- likely outflows -- in quasar--submm galaxy pairs \citep[100~kpc;][]{peng=2023}.
An in-depth analysis will be presented in Sun F.\ et al.\ (in prep.), but here we remark
that \darkhorse spectroscopy provides a compelling way to use faint emission-line
galaxies as a probe of diffuse CGM dust and, therefore, feedback mechanisms
in massive galaxies.

\subsection{Outflows}\label{s.sc.ss.outflows}

\jwst enabled the study of energy and momentum injection from star formation and
accretion at lower \mstar than what was possible before \citep{xu+2023,
carniani+2024,cooper+2025}. Among the unexpected discoveries is the high incidence rate of
neutral-gas outflows, traced primarily by \NaIall seen in absorption
\citep{davies+2024,belli+2024,deugenio+2024,taylor+2024,wu+2024,valentino+2025}.
These outflows seem to have much higher mass loading than the ionized phase
\citep[e.g.,][]{belli+2024,deugenio+2024}. Crucially, they are seen also in quiescent 
galaxies, including old systems \citep{sun+2025a}, suggesting a role in keeping these 
galaxies from forming new stars.

In \darkhorse, we report the discovery of multi-phase gas outflows in ID~175773 at
$z=2.54$. This dusty galaxy was selected with weight 10 (i.e. overriding the 300~nJy threshold from Section~\ref{s.tgt}). It hosts an AGN identified via a a tentative broad \Halpha component
(Section~\ref{s.sc.ss.blagn}) and BPT classification. Narrow-line detections include
hydrogen recombination lines \Hbeta, \Halpha, \Padelta, \Pagamma, and \Pabeta.
The forbidden lines \OIIIall and \SIIIall ~show a blueshifted wing, confirming the
presence of outflows previously detected using medium-band imaging \citep[][their source 209026]{zhu+2025}.

\begin{figure}
  {\phantomsubcaption\label{f.outflow.a}
   \phantomsubcaption\label{f.outflow.b}
   \phantomsubcaption\label{f.outflow.c}}
  \includegraphics[width=\columnwidth]{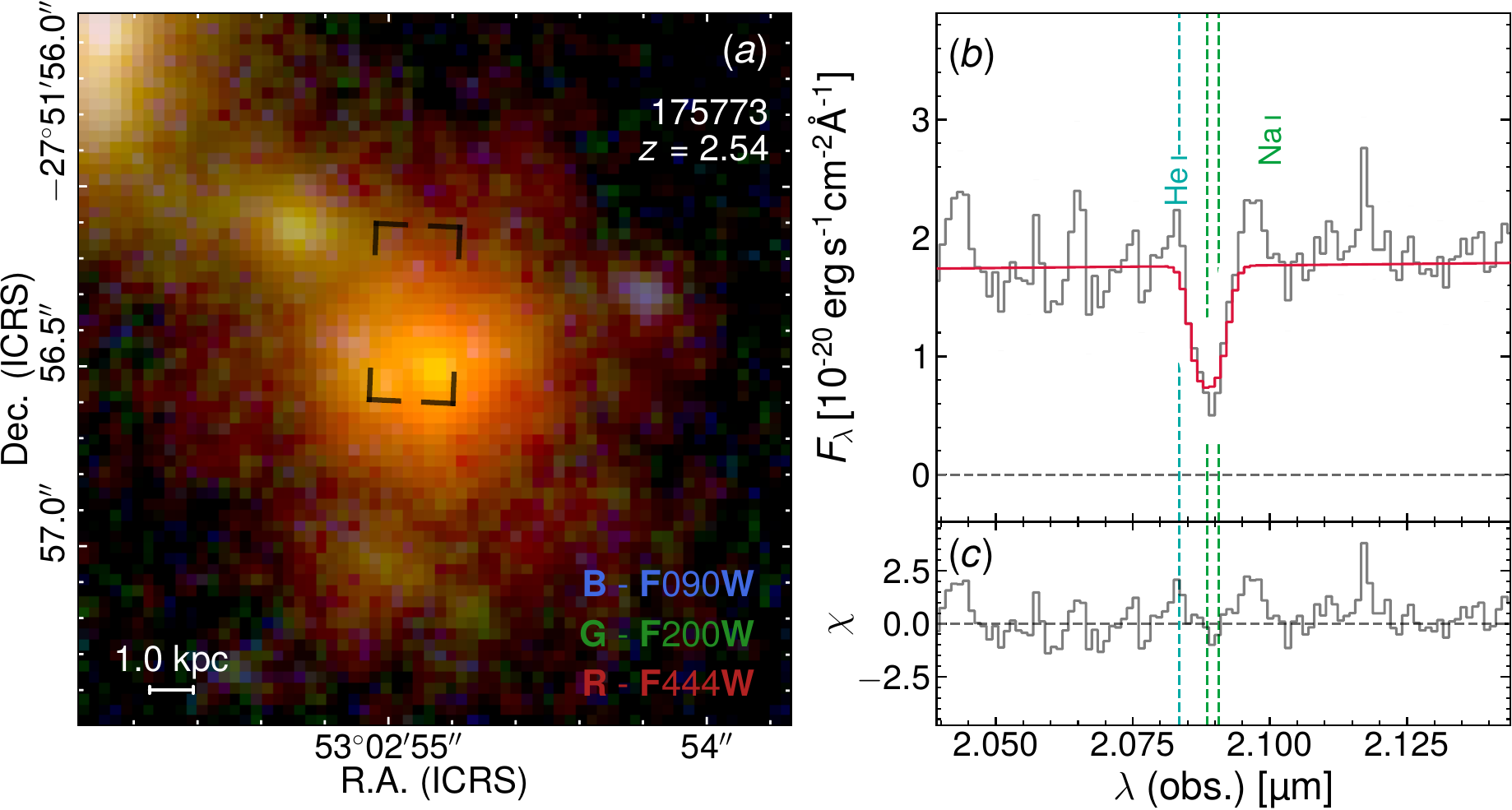}
  \caption{A dust-obscured AGN-host galaxy at $z=2.5$ (ID~175773), displaying compact morphology
  and red colour in NIRCam (panel~\subref{f.outflow.a}). \darkhorse reveals deep
  \NaIall absorption with a small (50~\kms) blueshift, tracing the galaxy ISM or
  a neutral-phase outflow
  (panels~\subref{f.outflow.b} and~\subref{f.outflow.c}). \darkhorse can take
  advantage of deep observations with high multiplex to also observe rare, bright
  sources for science cases that require long exposures.
  }\label{f.outflow}
\end{figure}

We detect clear absorption near \NaIall, blueshifted by 50~\kms. Our model
follows the approach of \citet[][see also \citealp{davies+2024,scholtz+2024,
perez-gonzalez+2025}]{rupke+2005}.
The model parameters are inferred in a Bayesian framework, using Markov Chain Monte
Carlo integration (MCMC), to fully explore the degeneracies between correlated
parameters \citetext{e.g., covering factor $C_f$ and sodium optical depth at line
centre, $\tau_0(\NaIL[5896])$; \citealp{davies+2024}}. The resulting fiducial model
is shown in Fig.~\ref{f.outflow}.
Using the geometry, metallicity and ionization assumptions of \citet{sun+2025a}, we estimate a mass outflow rate of
$\log(\dot{M}_\mathrm{out}/(\Msun\,\peryr)) = 2.0\pm0.2$. As a reference,
the maximal SFR (assuming no AGN contribution to the narrow component
of \Halpha) is $\log(\mathrm{SFR}/(\Msun~\peryr)) = 0.8\pm0.4$. So the mass loading factor
is clearly above unity, confirming that neutral-phase outflows are capable of shutting
down star formation \citep{belli+2024,deugenio+2024}.

Overall, these results show the potential for \darkhorse `hybrid' surveys. Whenever the science goals require deep spectroscopy of sources with low on-sky density -- and provided multi-band NIRCam coverage is available -- any remaining detector real estate can be allocated to overlapping faint-source spectroscopy.

\subsection{Mini-quenched Galaxies}\label{s.sc.ss.miniq}

In addition to exploring faint and rare emission-line objects, the survey speed and
depth of \darkhorse enable compelling non-detection science: the combination of
stringent upper limits on strong lines with the continuum shape from deep NIRCam
photometry enables to study star-formation burstiness which much
higher precision than what is possible from photometry alone -- even when using joint
constraints from medium- and wide-band observations \citep[e.g.,][]{tacchella+2022,simmonds+2025b}.

These improved constraints from fitting simultaneously photometry and spectroscopy can be used to study the stochasticity of SFR at high
redshifts. Deep prism observations have shown the existence of low-mass galaxies
($\mstar \lesssim 10^9~\Msun$) with negligible SFR \citep{strait+2023,looser+2024,baker+2025b}.
This population of `mini-quenched' galaxies are usually identified by their Balmer breaks
\citep[e.g.,][]{kuruvanthodi+2024,covelo-paz+2025}, lack of emission lines,
and UV slopes $\beta_\mathrm{UV}\sim-2$, indicating lack of current star formation but very young light-weighted age.
Their number abundance and physical properties hold compelling constraints on the
duty cycle of star formation and on the timescale and efficiency of feedback
\citep{gelli+2024,gelli+2025,dome+2024,dome+2025,mcclymont+2025,ceverino+2018,lovell+2023,
trussler+2025}.
The rapid cessation of star formation in these systems seems to require stronger
or more efficient feedback than just supernovae \citep{gelli+2024}, and may point
to bright quenched phases with little or no dust \citep{ferrara+2025,
baker+2025b}. While deep, medium-band photometry can identify `lulling' or
`smouldering' galaxies \citep{trussler+2025}, direct spectroscopic constraints
provide better time resolution than photometry alone, due to $\sim 100\times$
narrower effective bandwidth. The price to pay, of course, is completeness, which
\darkhorse spectroscopy mitigates substantially.

\begin{figure*}
  {\phantomsubcaption\label{f.mini.a}
   \phantomsubcaption\label{f.mini.b}
   \phantomsubcaption\label{f.mini.c}
   \phantomsubcaption\label{f.mini.d}}
  \includegraphics[width=\textwidth]{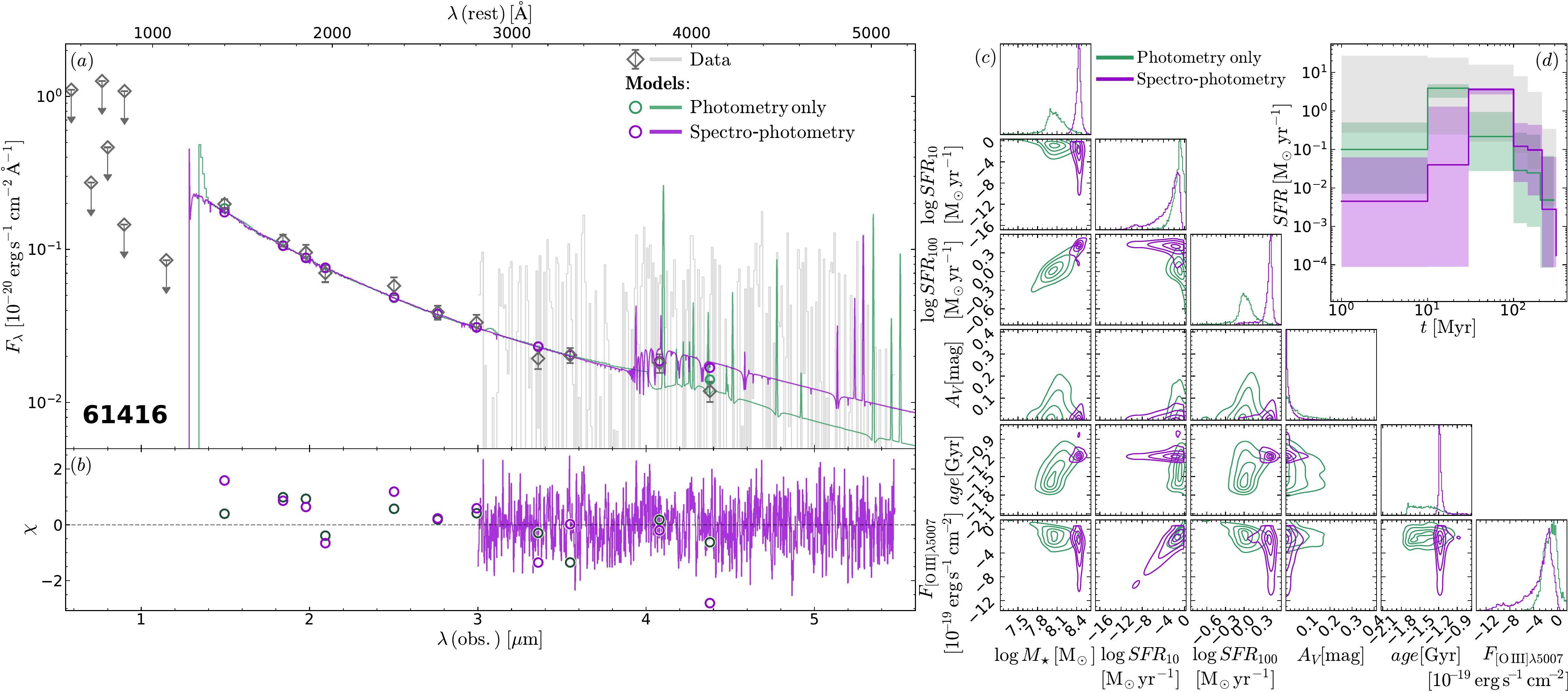}
  \caption{Including spectra from \darkhorse-undetected sources like 61416 at $z_\mathrm{phot}=9.6$ can significantly
  impact the resulting SED model. The green lines and circles show the fiducial
  \prospector model based on NIRCam photometry only, whereas the purple lines and
  circles are the model based on joint fitting of NIRCam and the NIRSpec G395M
  spectrum from the Pilot Survey. The photometry-only model predicts bright \OIIall
  that should be detectable by NIRSpec; therefore, including the observed spectrum
  rules out this solution and forces \prospector to seek a lower-SFR model, revealing
  61416 as a `mini-quenched' galaxy \citep[similar to ][]{looser+2024,baker+2025b}.}\label{f.mini}
\end{figure*}

To illustrate the improved sensitivity in the low-SFR regime, we report the
identification of a mini-quenched galaxy, ID~61416 at $z_\mathrm{phot}=9.6$
(Fig.~\ref{f.mini}).
We use photometry from the upcoming JADES Data Release 5 \citep{johnson+2026,robertson+2026},
and we compare the star formation histories obtained by fitting the photometry alone (indicated by the green lines and symbols in Fig.~\ref{f.mini}) vs a joint fit of the same photometry plus the \darkhorse spectrum (purple lines and symbols in Fig.~\ref{f.mini}).
We use
the Bayesian spectral energy distribution (SED) tool \prospector
\citep{johnson+2021}, with the same setup as \citet{deugenio+2025e}.
Specifically, we use a rising SFH prior, inspired by the success of models
of fixed baryon-conversion efficiency \citep[e.g.,][]{tacchella+2018}, and
implemented as described in \citet[][the rising prior is shown in grey
Fig.~\ref{f.mini.d}]{turner+2025}. We use a flexible dust-attenuation law
\citep{noll+2009,kriek+conroy2013}, with extra attenuation for stars younger than
10~Myr, representing dust in birth clouds \citep{charlot+fall2000}. The redshift
is free to vary, with a Gaussian prior set by the \eazy photometric redshift, and
truncated between $7.5\leq z \leq11$, dictated by the
photometric drops seen in NIRCam. The photometry and \darkhorse G395M spectrum
are shown in Fig.~\ref{f.mini.a}. We optimize the model twice: once based on
photometry only (green lines and circles in Fig.~\ref{f.mini}), and once based
on joint modelling of the photometry and \darkhorse spectroscopy (purple). The
fiducial (maximum \textit{a-posteriori}) models are shown in Fig.~\ref{f.mini.a},
with the residuals in panel~\ref{f.mini.b}. The redshift of the source is strongly
constrained by the non-detection in F115W; the photometry-only model interprets
the moderate F410M photometric excess as strong \OIIall emission, which should
be detectable by NIRSpec.
Lacking such detection in our data (grey spectrum in panel~\subref{f.mini.a}),
the spectro-photometric model revises the solution to an even lower recent SFR,
suggesting that this galaxy is a `mini-quenched' system, possibly the
highest-redshift such system \citep{strait+2023,looser+2024,baker+2025b}. The
blue UV slope and extremely low dust attenuation (Fig.~\ref{f.mini.c}) make
61416 an excellent candidate for a `blue monster' \citep{ziparo+2023,
ferrara+2025}.

\subsection{Broad-line AGNs}\label{s.sc.ss.blagn}

The Pilot Survey observed eight broad-line AGN (and one additional tentative source),
of which seven were targeted
(MSA weight 10; Table~\ref{t.blagn.summary}), and two were
weight 1. Upon inspection of their NIRCam SED,
two AGN (175773 and 171973) show weak UV emission (as expected from dust reddening) while the spectra show clear \NII emission; they appear to be dust-obscured AGN \citep[171793 is also an X-ray emitter;][]{kocekvsi+2025}. The other seven display the V-shaped SED typical of `Little Red Dots' \citep[LRDs;][]{matthee+2024,williams+2024}.
Thanks to
its straightforward mask allocation with two priority classes, we can weight these detections
by the inverse of the acceptance rates, which were 0.66 and 0.35 for weight 10 and
weight 1, respectively (Section~\ref{s.tgt.ss.tgt}). The completeness-corrected number
of broad-line AGN would then be $\approx 17$.

With the depth of the grating observations, \darkhorse can measure accurate broad-line 
shapes, line fluxes, and even auroral line fluxes. Of course, the current limitation is the
inability to measure accurate EWs, which would require forward modelling and removal
of all overlapping sources.

\begin{figure*}
  \includegraphics[width=\textwidth]{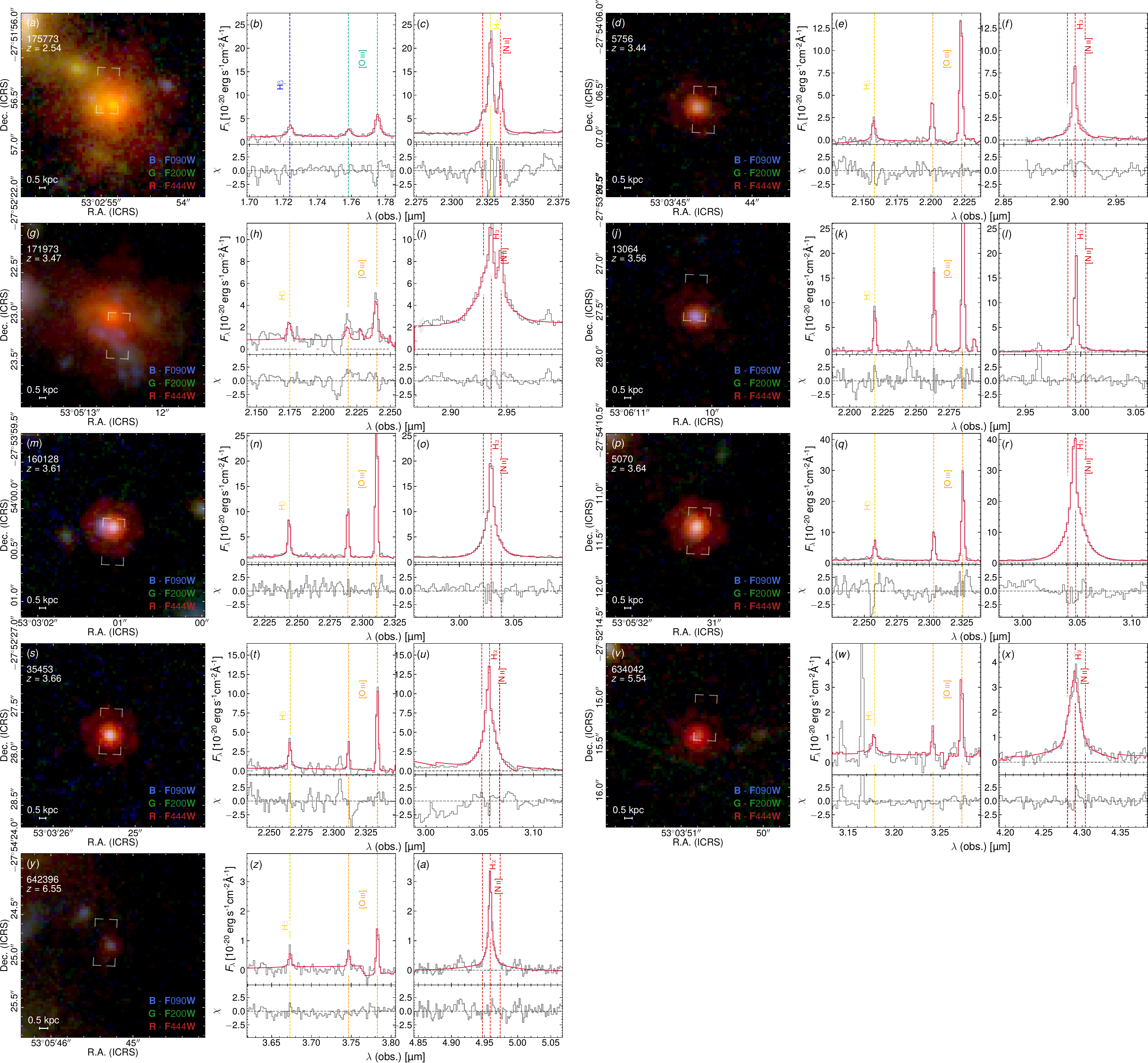}
\caption{A collection of broad-line AGN (and candidates) in \darkhorse, including seven LRDs.
For each target, we show the NIRCam RGB image, and the spectrum and fiducial model in the
region of \Hbeta--\OIIIall and \Halpha-\NIIall. \method can rapidly gather compelling samples
of high-quality LRD spectra with broad spectral coverage, enabling the study of the broad-line profiles, absorbers, narrow-line properties, and photo-ionization diagnostics. We find a remarkable range of broad-line shapes and luminosities.}\label{f.blagn.halpha}
\end{figure*}

To demonstrate the power of \darkhorse, we measure the narrow- and broad-line fluxes
of the seven AGNs, with the same approach as outlined in \citet{deugenio+2025e,deugenio+2025f}.
We mask prominent emission lines, then use a cubic spline to fit the continuum, with linear
interpolation across masked regions. The narrow lines are modelled as Gaussians with common
redshift and velocity dispersion; for doublets arising from the same upper level, we model
the two lines using the flux of the brightest line as free parameter, and the the fixed flux
ratio given by atomic physics, which we retrieve from \textsc{pyneb} \citep{luridiana+2015}.
For any doublet arising from the same lower level (\OIIall and \SIIall) the free parameters are
the flux of the bluest line and the doublet flux ratio, with the latter constrained to the range
allowed by atomic physics. We model only three narrow Balmer lines, \Halpha, \Hbeta and \Hgamma,
using fixed flux ratios appropriate for Case-B recombination, $\Telec=10,000$~K and $\nelec=100~\percm$.
The actual fluxes are then modulated using the \citetalias{gordon+2003} dust-extinction law,
parametrized by the attenuation \AV. For the LRDs, the broad lines are the sum of a Gaussian and an Gaussian convolved with an exponential kernel,
representing respectively direct light from the broad line, and light scattered by free electrons \citep{rusakov+2025}.
The scattering process is parametrized by the optical depth $\tau_\mathrm{e}$ and by \Telec
\citep[see][for more details]{deugenio+2025f}.
Finally, we also model \Halpha absorption,
using the same approach outlined in \citet{juodzbalis+2024b} and Section~\ref{s.sc.ss.outflows}.
A compilation of the broad-\Halpha lines
is shown in Fig.~\ref{f.blagn.halpha}, highlighting the diversity of profile shapes, narrow-to-broad flux ratios, and \OIIIL/\Hbeta ratios.

\begin{table}
\addtolength{\tabcolsep}{-0.5em}
	\centering
	\caption{Summary of broad-line AGNs (including LRDs) in \darkhorse.}
	\label{t.blagn.summary}
	\begin{tabular}{lllcccc}
		\hline
ID & R.A.     &    Dec.    & \!Weight\! & $z_\mathrm{phot}$$^a$ & $z_\mathrm{spec}$ & \!\!\!\!LRD \\
		\hline
175773$^b$ & 53.048578 & -27.865698 & 10 &  6.02 &  $2.5486\pm0.0001$   & 0 \\
  5756     & 53.062410 & -27.901838 & 10 &  3.72 &  $3.43725\pm0.00003$ & 1\\
171973$^c$ & 53.086836 & -27.873046 & 10 &  5.43 &  $3.4715\pm0.0002$   & 0 \\
13064      & 53.102832 & -27.890974 & 1  &  3.67 &  $3.56335\pm0.00002$ & 1 \\
160128     & 53.050301 & -27.900055 & 10 &  4.73 &  $3.61401\pm0.00002$ & 1 \\
  5070     & 53.092004 & -27.903136 & 10 &  3.62 &  $3.64372\pm0.00002$ & 1 \\
 35453     & 53.057029 & -27.874376 & 10 &  5.06 &  $3.65981\pm0.00001$ & 1 \\
634042     & 53.064113 & -27.870934 & 10 &  6.05 &  $5.5359\pm0.0005$   & 1 \\
642396     & 53.095910 & -27.906901 &  1 &  5.83 &  $6.5535\pm0.0002$   & 1 \\
		\hline
    \end{tabular}
    
\raggedright{Objects with `LRD' flag 1 have a V-shaped SED in NIRCam.
$^a$ We show photometric redshifts from the internal catalogue v0.9.5 RC2.

$^b$ Tentative broad line; classified as type 2 AGN using the BPT diagram

\citep{baldwin+1981}.
$^c$ X-ray emitter, see also \citet{kocekvsi+2025}.

}
\end{table}

\subsection{Overdensities and Clustering}\label{s.sc.ss.od&clust}

Since the high spectroscopic completeness and efficiency of this program are ideal for constructing spectroscopic redshift catalogues, this program is also well suited for identifying high-redshift galaxy overdensities and performing clustering analyses. Given the improved sensitivity of NIRSpec/MSA when compared to NIRCam/WFSS, fainter galaxies can be identified, thereby probing galaxies that are nearly an order of magnitude less luminous than galaxies selected by NIRCam/WFSS at fixed survey time.

\begin{figure*}
  \includegraphics[width=\textwidth]{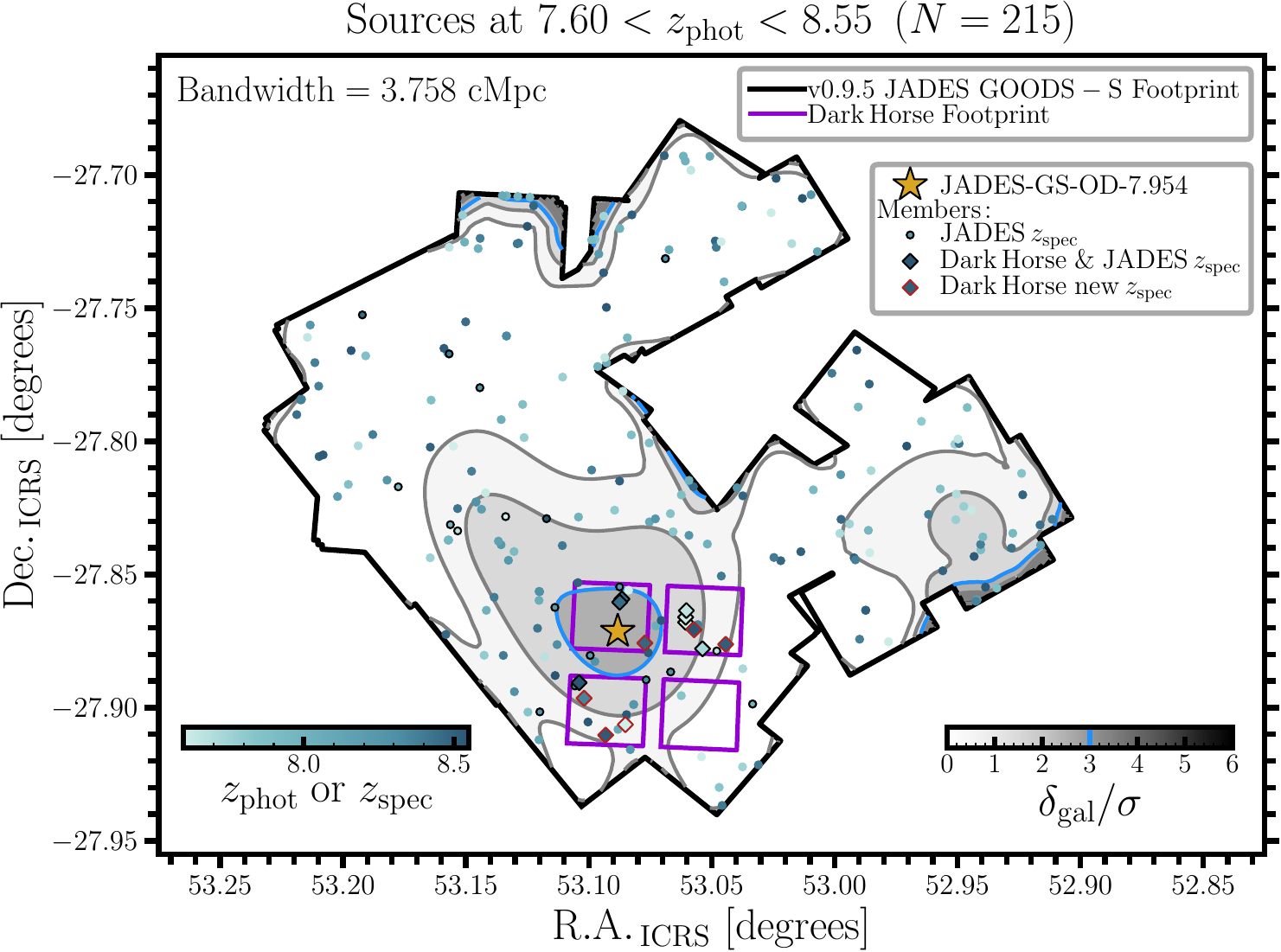}
    \caption{Dense-shutter NIRSpec/MSA is particularly well suited to study high-redshift galaxy environments by combining the improved completeness of slitless spectroscopy with the improved depth of slit spectroscopy. Here we show the spatial distribution of $N = 215$ galaxy candidates at $7.60 < z_{\mathrm{phot}} < 8.55$ from the JADES photometric sample, colour-coded by their photometric redshift, or spectroscopic redshift, if available. The thick black polygon represents the JADES/NIRCam GOODS-S footprint. The underlying density field of the photometric sample is determined using a KDE, following the methodology from \citet[][]{helton+2024,deugenio+2025c}. The contours illustrate the estimated density field and the blue contours represent a significance level of $3 \sigma$. Aside from the $> 3$~\textsigma peaks near the boundary of the footprint, which are the result of edge effects, the only other significant peak is spatially coincident with a spectroscopically confirmed galaxy overdensity at $z = 7.954$ \citep[JADES-GS-OD-$7.954$;][]{helton+2024}, as shown by the gold star. The \darkhorse single MSA pointing (purple outline) identifies $13$ spectroscopically confirmed members of this overdensity (diamonds with black outlines), including six new members (diamonds with red outlines).}\label{f.overd}
\end{figure*}

As a first look at the scientific prospects of using dense-shutter NIRSpec/MSA for overdensity identification and clustering analysis, we search for new members of an extreme galaxy overdensity at $z \approx 8$ identified by \citet{helton+2024}, JADES-GS-OD-$7.954$, which is the third highest redshift spectroscopically confirmed galaxy overdensity to date. The location of JADES-GS-OD-$7.954$ is illustrated by the gold star in Fig.~\ref{f.overd} and falls within the Dark Horse Pilot Survey pointing. This protocluster candidate originally contained four galaxies identified with NIRCam/WFSS from FRESCO and represented a density around $\sim 6$ times that of a random volume. Fig.~\ref{f.overd} also shows the spatial distribution of $N = 215$ galaxy candidates at $z \approx 8$ from the JADES photometric sample to further demonstrate the significance of the  overdensity JADES-GS-OD-$7.954$ \citep[including six \Lyalpha emitters;][]{jones+2025}. Following the methodology described in \citet[][]{deugenio+2025c}, we determine the underlying density field of the photometric sample by using a kernel density estimator (KDE). To quickly summarize, we assumed Gaussian kernels for the KDE and optimized the bandwidth (also known as the smoothing scale, as shown in the upper left corner of Fig.~\ref{f.overd}) by maximizing the likelihood cross-validation quantity. The estimated density field is illustrated by the contours, where purple contours represent overdensities above a significance level of $3 \sigma$. The only significant peak not affected by edge effects is spatially coincident with JADES-GS-OD-$7.954$.

It is obvious from looking at the upper panel of Fig.~\ref{f.zcomp} that there is an abundance of galaxies at $z \approx 8$. Indeed, 13/29 galaxies at $z > 7$ are within $\Delta z = 0.1$ of JADES-GS-OD-$7.954$, demonstrating the importance of cosmic variance when selecting samples of the highest redshift galaxies. Out of these $13$ galaxies, six are new discoveries, demonstrating the efficacy of using dense-shutter NIRSpec/MSA for identifying faint populations of clustered galaxies. Future work should look into identifying overdensities and voids with dense-shutter NIRSpec/MSA, especially given the recent discovery of pristine massive star formation caught at the break of cosmic dawn with NIRCam/WFSS in Abell~2744 \citep[][]{morishita+2025}.

\section{Discussion}\label{s.disc}

\subsection{Dark Horse relative to slitless spectroscopy}\label{s.disc.ss.vsgrism}

Having now presented the results of this programme, we turn to some more detailed aspects of the trade-off relative to the \jwst slitless grism modes.  We stress that while we see considerable advantages of the \method method, there are also ways in which the grism might be preferred, such that each method has preferred applications.

\subsubsection{Emission-line sensitivity}\label{s.disc.ss.vsgrism.sss.sens}

\begin{figure}
  {\phantomsubcaption\label{f.sensgrism.a}
   \phantomsubcaption\label{f.sensgrism.b}}
  \includegraphics[width=\columnwidth]{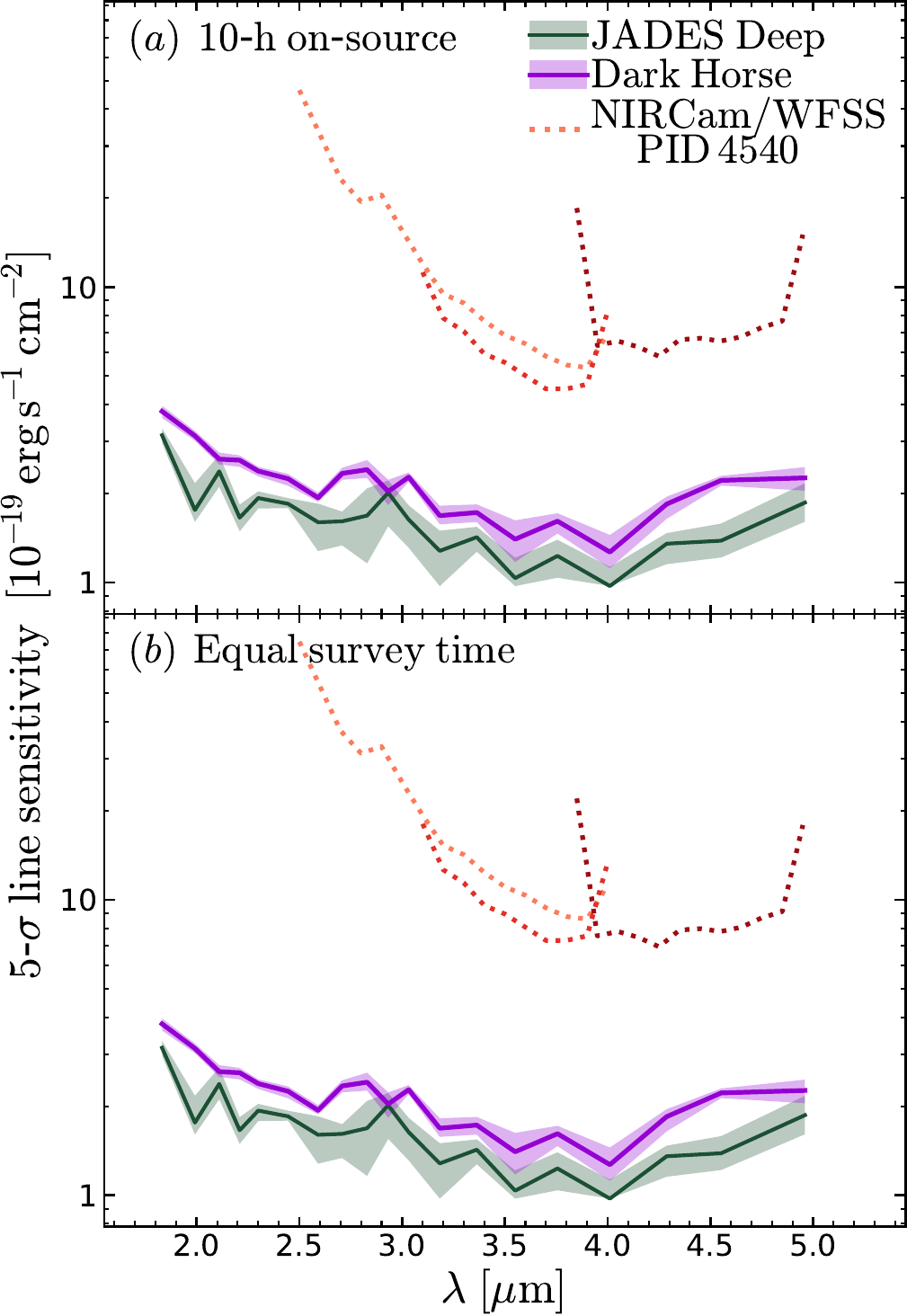}
\caption{A comparison between \method spectroscopy and NIRCam/WFSS. The purple and green lines are
the same as in Fig.~\ref{f.sens}, where we use nod background subtraction
(while master background would deliver even better sensitivity).
The red dotted lines show the median NIRCam/WFSS 5-\textsigma sensitivity with PID4540
in the F322W2, F356W, and F444W band (3--5.4-hour exposures; \citealp{eisenstein+2023b,sun+2026}).
In panel~\subref{f.sensgrism.a} we scale the sensitivity of NIRCam/WFSS to the same on-source time as NIRSpec/MSA,
while panel~\subref{f.sensgrism.b} shows the actual measured values, which correspond to approximately the same
survey time. Note that NIRCam/WFSS delivers 50~percent more sources with $z>3$ than \method.
}\label{f.sensgrism}
\end{figure}

To assess the \method method relative to slitless, we start with a comparison of their emission line sensitivities
(Fig.~\ref{f.sensgrism}), comparing to the deep NIRCam/WFSS observations from \jwst Cycle-3 PID~4540
(\citealt{eisenstein+2023b}; \citealp{sun+2026}.). This programme used the F322W, F356W and F444W filters and GRISMC to obtain grism spectroscopy of galaxies in the JOF across 2.4--5.0\,\mum, and the total grism exposure time is 82\,ks, comparable to that of \darkhorse (69.4~ks).
PID~4540 uses a $2\times2$ mosaics of NIRCam pointing, and therefore the median integration time is $\approx 3\text{--}5.4$ hours in each of the three bands. 
Improving the grism sensitivity to 10-hours on-source equivalent (matching the on-source time of \darkhorse),
the difference of depth between NIRCam/WFSS and \darkhorse is a factor of $\approx3$ (Fig.~\ref{f.sensgrism.a}).
Of course, this would require at least a 2.5 longer survey with NIRCam, given the actual survey times.

In contrast, in Fig.~\ref{f.sensgrism.b}, we show a comparison of the \textit{actual} sensitivities, achieved at
almost equal on-source time (actually, slightly longer for NIRCam). The median 5-\textsigma emission-line sensitivity
of PID~4540 for unresolved lines is $\approx5$ times of that for \darkhorse across 3.3--5\,\mum, a factor of 25 in
integration time. The sensitivity improvement is even larger at shorter wavelengths, exceeding a factor of 10 (100 in
integration) at $\lambda<3~\mum$.

This comparison highlights the high efficiency of the \method approach in obtaining deep emission-line spectroscopy
of a large sample of high-redshift galaxies. 
We also note that preliminary analyses of PID~4540 \citep{sun+2026} yield $\approx$50~percent more redshifts at $z>3$ than \darkhorse.
This is a result of the higher completeness with slitless spectroscopy and the flux density cap employed for \darkhorse.

\subsubsection{Geometry, field of view, and completeness}\label{s.disc.ss.vsgrism.sss.geom}

When it comes to survey speed, the instrument field of view matters. The total footprint of the four MSA quadrants of NIRSpec is about 9.82~arcmin$^2$, similar to the NIRCam field of view (9.30~arcmin$^2$) and about twice that of NIRISS.
It is important to note that the relatively high-resolution of the NIRCam grism means that the spectral traces are long and hence the coverage at any given wavelength is often limited by the field stop, reducing the field of view. In contrast, with the MSA gratings, one obtains the full wavelength coverage per target, save the small detector gap. 
However, single-configuration completeness greatly reduces the effective footprint of the MSA,
even when allowing spectral overlaps. Since NIRSpec picks from a catalogue, the actual penalty
depends on the target density of the parent sample (Section~\ref{s.disc.ss.samp}). Here, we
use the completeness of \darkhorse to evaluate the method, i.e. 854 out of 2,516 allocated targets,
or 35 percent (Section~\ref{s.tgt.ss.completeness}).
The dominant factors are MSA slit vignetting (0.64 with our choice of \texttt{entire\_open}) and the three-shutter requirement (0.60), with a smaller penalty due to exclusion zone in the dispersion direction (0.9). Thus the effective
MSA footprint is $9.82~\mathrm{arcmin}^2 \times 0.64 \times 0.60 \times 0.9 \approx 3.4$~arcmin$^2$, which we will use below.
If high completeness in a given field is needed, one could use multiple MSA configurations offset by a half-integer or even third-integer shutter spacing to access most targets. Alternatively, increasing the number
of allowed overlaps can also be achieved, both in the spatial direction (by moving from 3-shutter to 2- or 1-shutter slitlets) and/or in the spatial
direction (by overriding \textsc{mpt} minimum target separation of 4 shutters).
We note that slitless spectroscopy does not immediately yield full completeness either, because of overlap with brighter sources; one needs to use multiple dispersion angles.

Operationally, slitless spectroscopy does not require pre-imaging for target selection, although such imaging may be desired for line association and for the science application. It is the only option for pure parallel observing, and observations allowing many mechanism moves can produce compelling data sets of both imaging and spectroscopy in novel fields \citep[e.g.,][]{sun+2025b}.
Further, it can be awkward that the NIRSpec footprint of four quadrants in a near square is rather different than the NIRCam footprint: one not only needs pre-imaging for \method to be possible, but typically one needs a filled mosaic of at least 2$\times$2 pointings if one is to cover the
$3.1\times3.4$~arcmin$^2$ bounding box of the NIRSpec MSA.

\subsubsection{Sensitivity, bandwidth, and survey speed}\label{s.disc.ss.vsgrism.sss.speed}

As expected, strong suppression of the background by the MSA and the larger pixels of NIRSpec
(0.1~arcsec compared to 0.06~arcsec in NIRCam LW and NIRISS) sharply suppresses noise. Comparing to PID~4540, which represents a 20~percent longer time investment than \darkhorse, \method delivers a 4--10-fold
improvement in line sensitivities above NIRCam, depending on wavelength (Fig.~\ref{f.sensgrism}).
This is nominally a factor of 16--100 advantage in exposure time.
We note that this result is still using nod-based same-pixel background subtraction; we expect
that the data could also be reduced with non-local modelling, or `master background' subtraction,
which would suppress the noise further and avoid self-subtraction within larger galaxies.
This would make the case for single-shutter configurations even more compelling. Indeed,
one could measure accurate line fluxes without performing any background subtraction at the pipeline level, as
demonstrated in Fig.~\ref{f.noback}, showing that -- at least for emission-line science which is most relevant here -- \textit{single-shutter configurations are both
feasible and less noisy, even without master background.}

But further, the instantaneous bandwidth of NIRSpec is wider. The G395M grating alone covers 2.9--5.3~\mum, even more than the union of the NIRCam F356W and F444W filters.
If one wants full wavelength coverage, this is another factor of 2 over the grism.
Of course, NIRSpec also offers the G140M and G235M gratings. Using all three would give spectra from 0.7 to 5.3~\mum. Importantly, NIRISS only goes as red as 2.2~\mum, while the NIRCam grism has a zero-dispersion wavelength of 3.95~\mum and hence a sharply reduced field of view (and reduced first-order throughput) shortward of 3~\mum, ending at 2.4~\mum. This strongly favours the use of G235M to cover the 2--3~\mum wavelength range; notably this disperser will capture the strong \Halpha or \OIIIL lines for
$1.6<z<5.3$, which is a workhorse range for galaxy evolution studies with \jwst, capturing the epoch when $\sim50$ percent of the stellar mass in the Universe formed \citep[e.g.][]{madau+dickinson2014}, and the emergence of low-mass, environment-quenched galaxies \citep{donnari+2021a,donnari+2021b}.

To place these considerations on a more quantitative footing, we consider as a figure of merit
the integral over wavelength of area times inverse variance (5-\textsigma point source), per
unit time.
For the NIRCam grism, we use data from the pure-parallel programme SAPPHIRES
\citetext{Slitless Areal Pure Parallel HIgh-Redshift Emission Survey, PID~6434, PI E. Egami;
\citealp{sun+2025b}, their table~2}.
Adopting units of (arcmin$^2$) (\fluxcgs[-18])$^{-2}$ (\mum \, ks$^{-1}$), and assuming
$\Delta \lambda = 0.2$~\mum, we obtain metrics of 0.653, 0.677, 0.585, and 0.638 for
356W R, F356W C, F444W R, F444W C, respectively (i.e., around 0.64 on average). This includes an
average correction to take into account 20~percent more noise in module B than module A.

For \darkhorse, we consider the effective MSA quadrant area of 3.4 arcmin$^2$, as described
in Section~\ref{s.disc.ss.vsgrism.sss.geom}. For G395M, we integrate the sensitivity curve (Fig.~\ref{f.sens}) and obtain a metric
of 7.5, roughly ten times higher than NIRCam. This suggests that the \method method
is ten times more efficient than doing both F356W and F444W with the grism.

To summarize, the key argument for the \method mode is that the improvements in line sensitivity, field of view, and wavelength coverage far outweigh the factor of $\sim$2--3 reduction of single-configuration completeness.

\subsubsection{Caveats}\label{s.disc.ss.vsgrism.sss.cave}

Here we discuss some specific scenarios, and considerations that are difficult to quantify
and may depend on the specific application.
While NIRSpec delivers nearly the full wavelength range on all objects, save the detector gap,
grism slitless spectroscopy yields a complex distribution of wavelength coverage across the field, which does
not enter our metric.
While the proposed metric is useful for the question of finding a single bright line in as
many objects as possible, if one cares only about a single wavelength, the grism could switch
to a medium-band filter, boosting inverse variance by a factor of 2.5. In this case, the
useable field of view would also be larger; for example, F410M could get a metric of around
2.5, comparable to \method (i.e. 7.5).
On the other hand, the proposed metric does not capture the requirement to have coverage of
two well-separated lines. The grism loses further efficiency here, unless one is tiling fairly big areas.
Of course, the reported advantage of G395M becomes even stronger in G235M, because the grism area and throughput drops badly shortward of 3~\mum. 

But there are arguments and opportunities for the grism modes as well. To start, the grism modes can have lower operational overheads. In particular, if one only needs relatively shallow line sensitivities, short exposures with the grisms may suffice, plus one is getting nearly full targeting completeness in one exposure.  There is no value in taking NIRSpec exposures shorter than about 500 seconds, as the detector noise and MSA overheads will be severe. We note that at this time, \jwst operations require direct images when NIRCam and NIRISS are used as the prime instrument; these do add overheads not dissimilar to the MSA target acquisition sequence. Having high single-exposure completeness is particularly useful for applications like 
surveys of quasar environments 
\citep[e.g.,][]{kashino+2023,wang+2023}:
here one is primarily interested in a known wavelength and a small neighbourhood around the quasar, blunting some of the advantages of MSA spectroscopy.

Slitless spectroscopy is better for the study of line morphologies.  Because of the complicated pattern of the MSA vignetting and the coarser NIRSpec pixel scale, extracting morphologies from MSA spectroscopy requires very careful modelling. We note that the combination of measuring the line with both a grism and a direct image can yield compelling dynamics studies \citep[e.g.,][]{nelson+24,danhaive+2025}. In addition, slitless spectroscopy is also more reliable for the measurement of line fluxes, particularly from extended objects, because one is simply using photometric techniques in the dispersed images.  Flux calibration through the MSA slits requires detailed modelling of the slit vignetting even for cases of known morphological profiles, and of course deviations from the assumed profiles cannot be corrected at all, as one has not measured the light beyond the slit.

\subsection{Impact of sample selection}\label{s.disc.ss.samp}

The figure of merit reported above depends linearly on the effective survey footprint,
which in turn is linked to the completeness and to the characteristics of the input sample.
The quoted completeness of 0.35 and effective area of
3.1~arcmin$^2$ are specific of design choices of \darkhorse, not of the \method method.

To start, the choice of 3-shutter slitlets imposes a loss of efficiency over single-shutter
observations, because the fraction of operable shutters is 0.74, but the fraction of operable
three-shutter slitlets is only 0.60. This means \darkhorse could have allocated $\approx 1,100$
targets with no loss in sensitivity (actually, with a modest gain, since we would open only
1,100 shutters, against the actual 2,562). This change of strategy would raise the
completeness to 0.43. These numbers have been verified with \mpt, but for the actual \darkhorse implementation we adopted a more conservative approach.

The value of 0.43 is close to the maximum efficiency ceiling; one could consider relaxing
the shutter acceptance zone, but this would overly penalize compact sources, more so
at short wavelengths, where the telescope PSF is smaller
than the 0.2-arcsec bar shadows on the MSA.
On the other hand, while we noted that \darkhorse operates in the regime where completeness
is set by vignetting and shutter operability, this is directly linked to the target density
$\approx 150\text{--}350~\mathrm{arcmin}^{-2}$ in the two imaging areas. For instance, removing
our redshift cut, we are left with 12,000 sources inside the four MSA quadrants, reaching
target densities of 700--1700~arcmin$^{-2}$. Using a single shutter per source, \mpt 
can allocate $\sim5,000$ sources, but this exceeds the
threshold where overlapping background is becoming the dominant noise source. In this case, to
contain background noise (i.e. max 10 sources per row, Section~\ref{s.tgt.ss.overlaps}), one would want to
stay below $342\times10 \sim 3,500$ sources, so the completeness would be $3,500/12,000\approx0.3$. On the other hand, additional gains are possible
if one allows multiple visits; in this case, one could \textit{reduce} the
shutter acceptance area, which for each mask gives priority to sources
with good slit centration and, therefore, higher throughput.

\subsection{Other NIRSpec dispersers}\label{s.disc.ss.disp}

We note that below 2~\mum, the background increases again, rising to 45--60~nJy
per NIRSpec pixel (applicable to JOF in December). Since the photon rate per frequency 
scales as $\lambda^{-1}$, the rise in background at short wavelengths causes a sharp rise
in background noise, roughly $2\times$ between 2~\mum and 1~\mum. With a background noise
of 4~$\mathrm{e^{-}}$, G140M suffers from a much stiffer noise penalty than the other two 
gratings, such that an over-allocation of two sources would already cause the background 
noise to reach the detector noise. An over-allocation of $\approx 7.5$, as used here,
would imply a time penalty of 3--4, erasing most of the advantage of \method over
non-overlapping spectroscopy.

When considering the high-resolution gratings, the background is dispersed 3$\times$, allowing
the ceiling of 0.43 completeness to be reached even for the highest target densities. The trade-off
here is spectral truncation at the red end of quadrants 1 and 2 of the MSA.

We next consider the trade-offs between using this \method approach with the grating method and
using the NIRSpec prism dispersion mode. While the PRISM spectra should typically not be overlapped because they are background-limited and hence would increase the dominant noise source, the spectra are short enough that one can reach multiplex of nearly 200.  Given that the PRISM covers the same wavelength range as the union of all three gratings, the difference in multiplex is only modest.  However, per unit exposure time, the PRISM cannot achieve the line flux sensitivity of the gratings because the higher throughput of the prism cannot offset the increased noise from under-resolving the continuum and background. For the detection of emission lines, the gratings are likely to win this multiplexing assessment. A future analysis of overlapping prism spectroscopy will clarify the trade-offs.

Of course, achieving continuum spectroscopy even at modest SNR is scientifically important.  This provides the detection of Lyman breaks, indeed a faster route to a redshift for UV-bright galaxies at $z>9.5$ than the detection of rest-UV emission lines, and the measure of Balmer and 4,000-\AA breaks, important for the study of older stellar populations.  

On the flip side, the PRISM under-resolves the emission lines.  The gratings are needed to deblend certain close lines as well as to measure dynamics and outflows (Section~\ref{s.sc.ss.outflows}). And as one gets fainter than 100~nJy, continuum detections require very long exposures, such that the emission lines provide much of the science.  Indeed, the combination of grating redshifts and multi-color NIRCam imaging can support modelling of the continuum shape, such as for Balmer breaks (Section~\ref{s.sc.ss.miniq}). This is particularly relevant for young galaxies, where the continuum
is dominated by young, hot stars or even nebular emission, which have low-EW features that are
difficult to capture even with the prism.

\section{Summary and Conclusions}\label{s.conc}

In this article, we presented the \darkhorse spectroscopy survey using NIRSpec/MSA in
`\method' mode. Building upon the experience of previous studies that allowed
moderate overlaps, \darkhorse
allows for full spectral overlap on the NIRSpec detector, bridging the multiplex
of slitless spectroscopy and the sensitivity of slit observations.
To be effective, \method spectroscopy must build upon existing, pan-chromatic and spatially contiguous NIRCam imaging.
For this survey, we leverage the JADES Origins Field \citep{eisenstein+2023b}
and obtain a single pointing with G235M and G395M, focusing on $z>3$ extragalactic
science (Section~\ref{s.tgt}). The existing deep NIRCam coverage in legacy extragalactic
fields provides many opportunities for using \method spectroscopy.

After reducing the data using off-the-shelf methods, we confirm the feasibility of
the method.
Spectral overlaps cause additional background noise, which we do detect.
With the adopted setup of \darkhorse, we predict a noise penalty of 1.30,
corresponding to a time penalty of 1.7 to achieve the same
sensitivity as standard MSA (Section~\ref{s.tgt.ss.overlaps}).
These numbers are confirmed experimentally (Section~\ref{s.datared}),
meaning we are able to predict the MSA performance in \darkhorse, and use this
knowledge for survey design. As we argue below, a time penalty of 1.7 is perfectly acceptable.

We find a redshift
success rate of at least 0.65, comparable yet higher than grating surveys of similar
depth ($\sim0.52$). This result is likely driven by the high-quality photometric
redshifts available in GOODS-S, and especially in JOF, where there is extensive
and deep medium-band coverage. Nevertheless, an important component of the success
rate could be the high completeness, which enables targeting compelling sources without
having to resort to `filler' targets.

What makes the \method spectroscopy worth it, of course, is the five-fold increase in survey speed
relative to moderate-overlap grating surveys (Section~\ref{s.zflux}).
To demonstrate the method -- aside from measuring redshifts and line fluxes --
we present several science applications of \method spectroscopy (Section~\ref{s.sc}). These
range from classic
studies of emission lines as probes of star formation (Section~\ref{s.sc.ss.sfms})
and metallicity (Section~\ref{s.sc.ss.metals}), to studies of multi-phase gas
outflows (Section~\ref{s.sc.ss.outflows}) and broad-line AGNs 
(Section~\ref{s.sc.ss.blagn}). Sensitive, high-completeness observations also enable
studies of CGM dust at $z\sim3.5$ (Section~\ref{s.sc.ss.dust}), and even non-detections add value
to existing photometry, by delivering informative flux upper limits that improve the
constraints on photometric SFHs (Section~\ref{s.sc.ss.miniq}).

Reviewing the above makes clear why the pursuit of a general-purpose spectroscopy of a high-density sample of faint galaxies in established \jwst imaging fields is so well suited to this \method
grating mode.  These sources often have strong emission lines that are both efficient for redshift determination and essential for physical characterization. The density of sources is very high, enough to support the high multiplex of this MSA mode.

While our \darkhorse survey identified no unmitigatable problems, further
optimizations are also possible, since there are trade-offs between different
MSA configurations and targeting choices (Section~\ref{s.disc}). These future
improvements could possibly make \method an even more compelling path to
efficiently obtain large numbers of \jwst spectra in deep cosmological fields,
with depths that cannot be reached by the grism or by any current ground-based instrument.

\section*{Acknowledgements}

We thank the anonymous referee for their insightful comments which helped clarify and improve this article.
This work is based on observations made with the NASA/ESA/CSA James Webb Space Telescope. The data were obtained from the Mikulski Archive for Space Telescopes at the Space Telescope Science Institute, which is operated by the Association of Universities for Research in Astronomy, Inc., under NASA contract NAS 5-03127 for JWST. These observations are associated with program \#3215.

FDE, RM, XJ, JS, IJ and GCJ acknowledge support by the Science and Technology Facilities Council (STFC), by the ERC through Advanced Grant 695671 ``QUENCH'', and by the
UKRI Frontier Research grant RISEandFALL.
DJE, CNAW, BDJ, BER, and ZJ acknowledge support from the NIRCam Science Team contract to the University of Arizona, NAS5-02015. DJE, JHM and BER also acknowledge support from JWST Program 3215. Support for program \#3215 was provided by NASA through a grant from the Space Telescope Science Institute, which is operated by the Association of Universities for Research in Astronomy, Inc., under NASA contract NAS 5-03127. DJE is supported as a Simons Investigator.
RM also acknowledges funding from a research professorship from the Royal Society.
SC and EP acknowledge support by European Union's HE ERC Starting Grant No. 101040227 - WINGS.
AJB, AJC and JC acknowledge funding from the ``FirstGalaxies'' Advanced Grant from the European Research Council (ERC) under the European Union's Horizon 2020 research and innovation program (Grant agreement No. 789056).
IJ also acknowledges support by the Huo Family Foundation through a P.C. Ho PhD Studentship.
ST acknowledges support by the Royal Society Research Grant G125142.
SA and MP acknowledges grant PID2021-127718NB-I00 funded by the Spanish Ministry of Science and Innovation/State Agency of Research (MICIN/AEI/ 10.13039/501100011033). MP also acknowledges the grant RYC2023-044853-I, funded by  MICIU/AEI/10.13039/501100011033 and European Social Fund Plus (FSE+).
ECL acknowledges support of an STFC Webb Fellowship (ST/W001438/1).
ALD thanks the University of Cambridge Harding Distinguished Postgraduate Scholars Programme and Technology Facilities Council (STFC) Center for Doctoral Training (CDT) in Data intensive science at the University of Cambridge (STFC grant number 2742605) for a PhD studentship.
YI is supported by JSPS KAKENHI Grant No. 24KJ0202.
The research of CCW is supported by NOIRLab, which is managed by the Association of Universities for Research in Astronomy (AURA) under a cooperative agreement with the National Science Foundation.
JW gratefully acknowledges support from the Cosmic Dawn Center through the DAWN Fellowship. The Cosmic Dawn Center (DAWN) is funded by the Danish National Research Foundation under grant No. 140.
The authors acknowledge use of the lux supercomputer at UC Santa Cruz, funded by NSF MRI grant AST 1828315.

This work made extensive use of the freely available \href{http://www.debian.org}{Debian GNU/Linux} operative system.
We used the \href{http://www.python.org}{Python} programming language \citep{vanrossum1995}, maintained and distributed by the Python Software Foundation. We made direct use of Python packages
{\sc \href{https://pypi.org/project/astropy/}{astropy}} \citep{astropy+2013,astropy+2018},
{\sc \href{https://pypi.org/project/corner/}{corner}} \citep{foreman-mackey2016},
{\sc \href{https://pypi.org/project/emcee/}{emcee}} \citep{foreman-mackey+2013},
{\sc \href{https://pypi.org/project/jwst/}{jwst}} \citep{alvesdeoliveira+2018},
{\sc \href{https://pypi.org/project/matplotlib/}{matplotlib}} \citep{hunter2007},
{\sc \href{https://pypi.org/project/numpy/}{numpy}} \citep{harris+2020},
{\sc \href{https://pypi.org/project/astro-prospector/}{prospector}} \citep{johnson+2021} \href{https://github.com/bd-j/prospector}{v2.0},
{\sc \href{https://pypi.org/project/PyNeb/}{pyneb}} \citep{luridiana+2015},
{\sc \href{https://pypi.org/project/python-fsps/}{python-fsps}} \citep{johnson_pyfsps_2023},
{\sc \href{https://pypi.org/project/pysersic/}{pysersic}} \citep{pasha+miller2023},
{\sc \href{https://github.com/honzascholtz/qubespec/}{qubespec}} \citep{scholtz+2025},
and {\sc \href{https://pypi.org/project/scipy/}{scipy}} \citep{jones+2001}.
We also used the softwares {\sc \href{https://github.com/cconroy20/fsps}{fsps}} \citep{conroy+2009,conroy_gunn_2010}, {\sc \href{https://www.star.bris.ac.uk/~mbt/topcat/}{topcat}}, \citep{taylor2005}, {\sc \href{https://github.com/ryanhausen/fitsmap}{fitsmap}} \citep{hausen+robertson2022} and {\sc \href{https://sites.google.com/cfa.harvard.edu/saoimageds9}{ds9}} \citep{joye+mandel2003}.

\section*{Data Availability}

This work is based on observations made with the NASA/ESA/CSA James Webb Space Telescope. Raw data were obtained from the \href{https://mast.stsci.edu/portal/Mashup/Clients/Mast/Portal.html}{Mikulski Archive for Space Telescopes} at the Space Telescope Science Institute, which is operated by the Association of Universities for Research in Astronomy, Inc., under NASA contract NAS 5-03127 for \jwst. These observations are associated with programme PID~3215.
Reduced spectra and catalogues are available on the JADES website 
\href{https://jades-survey.github.io/scientists/data.html}{https://jades-survey.github.io/scientists/data.html} and on the online database \href{https://jades.herts.ac.uk/DR4/DarkHorse}{https://jades.herts.ac.uk/DR4/DarkHorse}; for the data structure, please refer to the JADES DR4 \citep{scholtz+2025}.
Supporting imaging and catalogues from PIDs 1210, 1286, 1287, 1895, 1963, 3215, and 6434 are available through the JADES DR5 \citep{johnson+2026,robertson+2026}.
Supporting spectroscopic data from PID 1210 is available through the JADES public data release 1 \citep{bunker+2024}.
Supporting spectroscopic data from PIDs 1180, 1181, 1286, 1287, and 3215 is available through the JADES public data relases 3 and 4 \citep{deugenio+2025a,scholtz+2025}.

\section*{Affiliations}
\noindent
{\it
\hypertarget{aff11}{$^{11}$}Centro de Astrobiolog\'ia (CAB), CSIC–INTA, Cra. de Ajalvir Km.~4, 28850 - Torrej\'on de Ardoz, Madrid, Spain\\
\hypertarget{aff12}{$^{12}$}Institut d'Astrophysique de Paris, Paris, 98 bis Boulevard Arago, 75014 Paris, France\\
\hypertarget{aff13}{$^{13}$}Centre for Astrophysics Research, Department of Physics, Astronomy and Mathematics, University of Hertfordshire, Hatfield AL10 9AB, UK\\
\hypertarget{aff14}{$^{14}$}Department of Astronomy and Astrophysics University of California, Santa Cruz, 1156 High Street, Santa Cruz CA 96054, USA\\
\hypertarget{aff15}{$^{15}$}NSF National Optical-Infrared Astronomy Research Laboratory, 950 North Cherry Avenue, Tucson, AZ 85719, USA\\
\hypertarget{aff16}{$^{16}$}NRC Herzberg, 5071 West Saanich Rd, Victoria, BC V9E 2E7, Canada\\
\hypertarget{aff17}{$^{17}$}DARK, Niels Bohr Institute, University of Copenhagen, Jagtvej 155A, DK-2200 Copenhagen, Denmark\\
\hypertarget{aff18}{$^{18}$}AURA for European Space Agency, Space Telescope Science Institute, 3700 San Martin Drive. Baltimore, MD, 21210\\
\hypertarget{aff19}{$^{19}$}Instituto de Astrof\'isica, Pontificia Universidad Cat\'olica de Chile, Av. Vicuña Mackenna 4860, 782-0436 Macul, Santiago, Chile\\
\hypertarget{aff20}{$^{20}$}Max-Planck-Institut f\"ur extraterrestrische Physik (MPE), Gie{\ss}enbachstra{\ss}e 1, 85748 Garching, Germany\\
\hypertarget{aff21}{$^{21}$}Cosmic Dawn Center (DAWN), Copenhagen, Denmark\\
\hypertarget{aff22}{$^{22}$}Niels Bohr Institute, University of Copenhagen, Jagtvej 128, DK-2200, Copenhagen, Denmark
}



\bibliographystyle{config/mnras}
\bibliography{dhorse}




\bsp	
\label{lastpage}
\end{document}